\documentclass[12pt, draftclsnofoot, onecolumn]{IEEEtran}
\usepackage[compress]{cite}
\usepackage{graphicx}
\usepackage{color}
\usepackage{amsmath}
\usepackage{caption}
\usepackage{amssymb}
\usepackage{amsfonts}
\usepackage{psfrag}
\usepackage{booktabs}
\usepackage{array}
\usepackage[named]{algo}
\usepackage{algorithmic}
\usepackage{algorithm}
\usepackage{booktabs}
\usepackage{algorithm,setspace}
\IEEEoverridecommandlockouts
\usepackage[font=footnotesize]{subcaption}
\usepackage{mathtools}
\DeclareMathOperator*{\argmax}{arg\,max}

\makeatletter
\DeclarePairedDelimiter{\norm}{\lVert}{\rVert}

\newcommand{\Rmnum}[1]{\expandafter\@slowromancap\romannumeral #1@}
\makeatother

\begin{document}
\title{A VCSEL Array Transmission System with Novel Beam Activation Mechanisms}

\author{\IEEEauthorblockN{Zhihong Zeng, Mohammad Dehghani Soltani, Majid Safari and Harald Haas}\\
\thanks{\indent 
The authors are with the
\mbox{University} of \mbox{Edinburgh}, LiFi R\&D Centre, \mbox{Edinburgh}, EH9 3JL, UK, (e--mail: \{zhihong.zeng, m.dehghani, Majid.Safari, h.haas\}@ed.ac.uk.) This publication was made possible by the funding from PureLiFi Ltd. The statements made herein are solely the responsibility of the author[s]. Professor Harald Haas acknowledges support from the EPSRC under Established Career Fellowship Grant EP/R007101/1. He also acknowledges the financial support of his research by the Wolfson Foundation and the Royal Society. } 
}

\maketitle
\vspace{-30pt}

\begin{abstract}
Optical wireless communication (OWC) is considered to be a promising technology which will alleviate traffic burden caused by the increasing number of mobile devices.  In this study, a novel vertical-cavity surface-emitting laser (VCSEL) array is proposed for indoor OWC systems. To activate the best beam for a mobile user, two beam activation methods are proposed for the system. The method based on a corner-cube retroreflector (CCR) provides very low latency and allows real-time activation for high-speed users. The other method uses the omnidirectional transmitter (ODTx). The ODTx can serve the purpose of uplink transmission and beam activation simultaneously. Moreover, systems with ODTx are very robust to the random orientation of a user equipment (UE). System level analyses are carried out for the proposed VCSEL array system. For a single user scenario, the probability density function (PDF) of the signal-to-noise ratio (SNR) for the central beam of the VCSEL array system can be approximated as a uniform distribution. In addition, the average data rate of the central beam and its upper bound are given analytically and verified by Monte-Carlo simulations. For a multi-user scenario, an analytical upper bound for the average data rate is given. The effects of the cell size and the full width at half maximum (FWHM) angle on the system performance are studied. The results show that the system with a FWHM angle of $4^\circ$ outperforms the others.
\end{abstract}


\section{Introduction}
\label{SectionIntroduction}
The sixth-generation (6G) communication is required to support a variety of services such as the Internet of things (IoT), augmented reality, virtual reality, video streaming and online gaming with extremely high data and low latency. However, the current sub-6 GHz band is already congested, and lower-band communication is hard to support the high data rate transmission. The optical wireless communication (OWC) is a promising solution to support the extremely high rate transmission in a new band.  There are two main indoor OWC technologies: visible light communication (VLC) and the beam-steered infrared light communication (BS-ILC) \cite{Koonen_JLT_2018}. The VLC systems use wide-spread beams emerging from illumination systems while the BS-ILC systems use narrow well-directed beams emerging from a dedicated source. 

The VLC systems uses the visible light spectrum of 400 to 700 nm, which provides a bandwidth greater than 320 THz. Light-fidelity (LiFi) is a promising and novel bidirectional, high-speed  and  fully networked  VLC  technology \cite{LiFiHaas,MaShuai1}. By using orthogonal frequency division multiplexing (OFDM), a single light emitting diodes (LEDs) can achieve a data rate of $10$ Gbit/s \cite{Islim}. In addition, LiFi is able to provide data rates of $15.73$ Gbit/s using off-the shelf LEDs \cite{Bian:19}. Due to the wide coverage of the LED, multiple user equipment (UE) may be served by the same LED and may share the same resources, which may lead to traffic congestion when the load is high. In \cite{ZheADT}, an angle diversity transmitter (ADT) which consists of multiple narrow-beam visible light LEDs has been proposed for optical wireless networks with the space division multiple access scheme (SDMA). Multiple users at different locations can be served simultaneously by activating different LED elements on an ADT. The result shows that SDMA schemes can mitigate  co-channel interference and thus greatly increase the average spectral efficiency (ASE). However, the VLC systems requires the illumination to be switched-on, which may not always be desired and thus increase the power consumption. Moreover, typically, the white light emitting LED, which is used for illumination, have limited bandwidth as it is based on a blue LED with phosphor coating \cite{Koonen_JLT_2018}.  

The deployment of narrow infrared beams in BS-ILC systems leads to small cell coverage, which implies that it is more likely that each beam only serves a single UE. Therefore, capacity sharing among multiple UEs as well as traffic congestion can be avoided. By only activating beams that point towards the UEs, the BS-ILC systems provide better energy-efficiency as narrow beams with high directivity can send a greater portion of the transmitted power to the corresponding UE. Therefore, a higher signal-to-noise ratio (SNR) can be achieved. Moreover, due to the high bandwidth of narrow beam sources, such as optical fiber, laser diode and vertical-cavity surface-emitting laser (VCSEL), the link can support higher capacity. In addition, compared with wide-spread beams, the narrow beams provide better privacy as UEs outside the coverage area cannot receive the transmitted signals. The BS-ILC systems have been explored in \cite{Koonen_OECC_2014,Koonen_JLT_2016,Koonen_MWP_2017,Koonen_JLT_2018}. Wavelength-controlled 2D beam-steered systems based on fully-passive crossed-grating devices are introduced and 1D beam steering has been demonstrated in \cite{Koonen_OECC_2014}, which shows a multi-beam system with a capacity of 2.5 Gbit/s. The 2D steering of the multi-beam system has been first demonstrated in \cite{Koonen_JLT_2016} and with adaptive discrete multitone modulation (DMT) using 512 tones, a gross bit rate of 42.8 Gbit/s has been achieved. In addition, by using 60 GHz radio signal in the uplink, upstream delivery of 10 Gbit/s per upstream has been presented. In \cite{Koonen_MWP_2017}, the authors proposed a novel OWC receiver concept, which can enlarge the receiver aperture without reducing the bandwidth, Also, a multi-beam system with downstream capacities of up to 112 Gbit/s per infrared beam was demonstrated. In \cite{Koonen_JLT_2018}, an alternative approach, which is based on a high port count arrayed waveguide grating router (AWGR) and a high-speed lens, is proposed. With pulse amplitude modulation (PAM)-4 modulation, a total system throughput beyond 8.9 Gbit/s over 2.5 m can be achieved by using 80-ports C-band AWGR. Moreover, the localization and tracking of the UE are achieved based on a 60 GHz ultra-wideband (UWB) radio link. However, all these studies consider fixed user location without mobility and the latency is high in these systems. When a mobile UE is considered, due to the small confined coverage area of the narrow beam, the beam that serves a UE may vary fast and frequently. Therefore, a high-accuracy and low-latency beam activation system or user tracking system is required to achieve a seamless connection. 

Ordinary radio based indoor positioning techniques based on WiFi, Bluetooth, UWB and radio frequency identification (RFID) are not suitable candidates in this study due to their low accuracy, high latency and high hardware cost \cite{VLP_zhechen}.
In recent years, visible light positioning (VLP) has emerged and various algorithms for VLP systems have been proposed, which include received signal strength (RSS), time differential of arrival (TDOA) and angle of arrival (AOA) methods. Due to being able to achieve high positioning accuracy and low cost, VLP is attracting more and more attention. Some studies have shown that the VLP technologies can achieve centimeter-level accuracy \cite{VLP_zhechen,VLP_xu_2018,VLP_wenyuan_2020,HassanNaveed2015IPVLC_ACM,Mashuai2}, which is way more accurate compared to  Bluetooth (2-5 m), WiFi (1-7 m) and other technologies \cite{ZhuangSurveyPositioning}. The VLP systems can be divided into two categories based on the receiving device: photodiode (PD) based VLP or image sensor (IS) based VLP . The PD-based VLP system has low latency but is sensitive to the device rotation, and hence cannot achieve high positioning accuracy for UEs without fixed orientation. The IS-based VLP systems are more robust to device rotation. However, due to the high computational latency of image processing or the communication latency of transmitting image data for server-assisted computation, the real-time performance of these systems are limited \cite{VLP_zhechen}.  

Also, when VLP is adopted, the UE is required to autonomously transmit location information to the access points (APs) owing to the lack of a real-time backward channel from UE to APs. This adds an unnecessary burden on the UE and leads to the inevitable latency in real-time tracking.  As the VLP system requires illumination devices and real-time backward channels, it may not be suitable for the VCSEL array system proposed in this study. To address the issue of power consumption and latency caused by UE localization, two beam activation schemes are proposed in this study. The first method is a passive beam activation scheme using a corner-cube reflector (CCR). The CCR is a light-weight small device, which can reflect light back to its source with minimal scattering. By using CCR, immediate feedback can be obtained for beam activation which minimizes the latency and leads to almost zero delay. The CCR has been proposed for the VLP system in \cite{ShaoRetro2018}.  Compared with the method in \cite{ShaoRetro2018}, no illumination equipment is required in our method and the power consumption for the localization system is almost zero. The second method utilizes the omnidirectional transmitter (ODTx) in the uplink communication. By using the uplink RSS matrix, this method does not require any extra signal power for localization. Compared with the PD-based VLP system, the beam activation system with ODTx is robust to the device orientation, that is to say the user can be located accurately without concerning the random orientation of the UE. Compared with the IS-based system, image sensor is not required, which reduces the cost. And the computation latency is reduced as the positioning algorithm is based on the RSS, which is the simplest and most cost effective schemes.

Among different types of laser diodes, vertical cavity surface emitting lasers (VCSELs) are one of the promising candidates to ensure high-data rate communications due to several outstanding features such as \cite{Iga2000JSTQE}: high-speed modulation (bandwidths of larger than $10$ GHz), high power conversion efficiency, low cost and compact in size. These attributes make VCSELs appealing to many applications, particularly for high-speed indoor networks \cite{Liu19VCSEL}. In this study, a VCSEL array systems with novel beam activation methods are proposed. Compared with the wavelength-controlled BS-ILC systems, the proposed VCSEL array system removes the requirement of wavelength-tuning, spatial light modulation (SLM), microelectromechanical systems (MEMS) and coherence of beams and fiber connection to the AP. End-point devices may include virtual reality  devices, smartphones, televisions, computers and IoT applications. 
In summary, the main contributions of this work are listed as follows:
\begin{itemize}
\item We propose a VCSEL array system which can support high data rate, low latency and multiple UEs without the requirement of expensive/complex hardware, such as SLM, MEMS, fiber and so on. 

\item Two beam activation methods are proposed based on the small cell property of the VCSEL array system. The beam activation based on the CCR can achieve low power consumption and almost-zero delay, allowing real-time beam activation for high-speed users. The other beam activation is based on the ODTx, which serves the purpose of the uplink transmission and beam activation simultaneously. By collecting the RSS values, an artificial neural network (ANN) is trained to predict the index of serving beam directly without estimating the UE position first. This method is robust against random device orientation and is suitable for low-speed users.

\item For a single UE scenario, regarding the central beam, the probability density function (PDF) of the SNR is derived. The analytical derivation for the average data rate is provided for the central beam and an upper bound is presented for the VCSEL array system.

\item In terms of scenarios with multiple users, the optical SDMA is adopted and an analytical upper bound for the average data rate is developed.
 
\item The effects of the cell size and beam divergence angle are considered in this study. By evaluating the system performance, the choices of cell size and beam divergence angle are proposed for the VCSEL array system.
\end{itemize}

The rest of the paper is organized as follows. The VCSEL array system model is introduced in Section \ref{SectionSystemModel}. In Section \ref{SectionBeamActivation}, two beam activation methods are proposed. Analytical derivations for the system level performance are presented in Section \ref{SectionSystemLevelAnalysis}. The performance evaluation and discussion are presented in Section \ref{SectionSimulationResults}. Conclusions are drawn in Section \ref{SectionConclusion}.  

\section{System Model}
\label{SectionSystemModel}

\begin{figure*}[h!]
 \centering
 \includegraphics[width=0.5\textwidth]{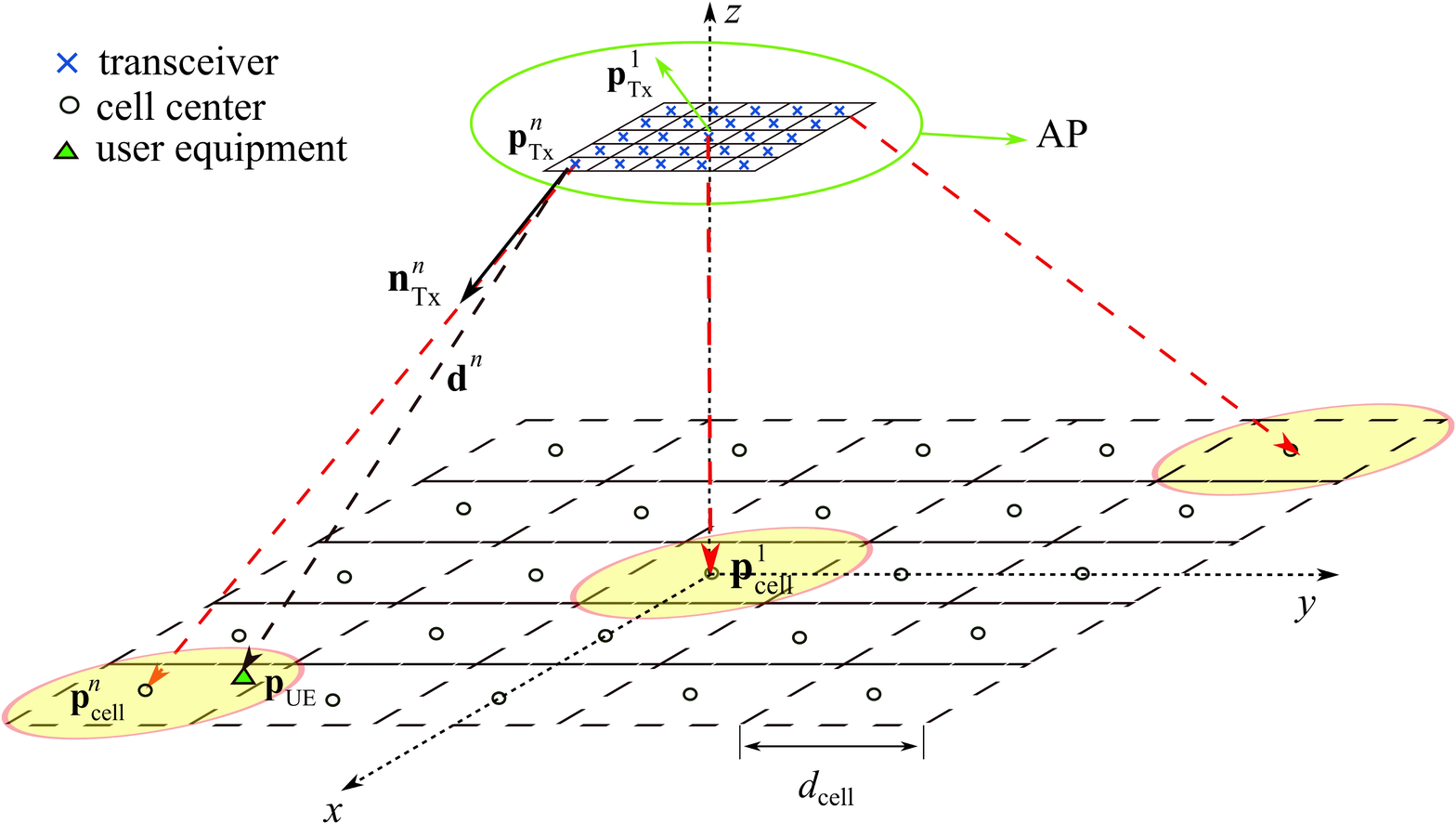}
\caption{The downlink geometry of a VCSEL array system.}
\label{fig_setup}
\end{figure*}

A VCSEL array system is presented in Fig. \ref{fig_setup}. The AP of the system is composed of $N_{\rm{beam}}$ narrow-beam Txs and each Tx is a VCSEL with a beam divergence of $\theta_{\rm{beam}}$. The position of the $n$-th Tx is denoted as ${\bf{p}}^{n}_{{\rm{tx}}}$. The $n$-th Tx is slightly tilted so that it is directed towards the center of its corresponding cell, ${\bf{p}}^{n}_{{\rm{cell}}}
$. Hence, the normal vector of the $n$-th Tx is denoted as ${\bf{n}}^{n}_{{\rm{tx}}}=({\bf{p}}^{n}_{{\rm{cell}}}-{\bf{p}}^{n}_{{\rm{tx}}})/\norm{{\bf{p}}^{n}_{{\rm{cell}}}-{\bf{p}}^{n}_{{\rm{tx}}}}$, where $\norm{\cdot}$ denotes the norm operator. The side length of each cell is denoted as $d_{\rm{cell}}$. The location of the UE is denoted as $\bf{p}_{\rm{UE}}$ and for the downlink communication, an avalanche photodiode (APD) is used as the receiver (Rx) at the UE side. 

\subsection{Gaussian Beam}
\begin{figure}[ht]
 \centering
 \includegraphics[width=0.5\textwidth]{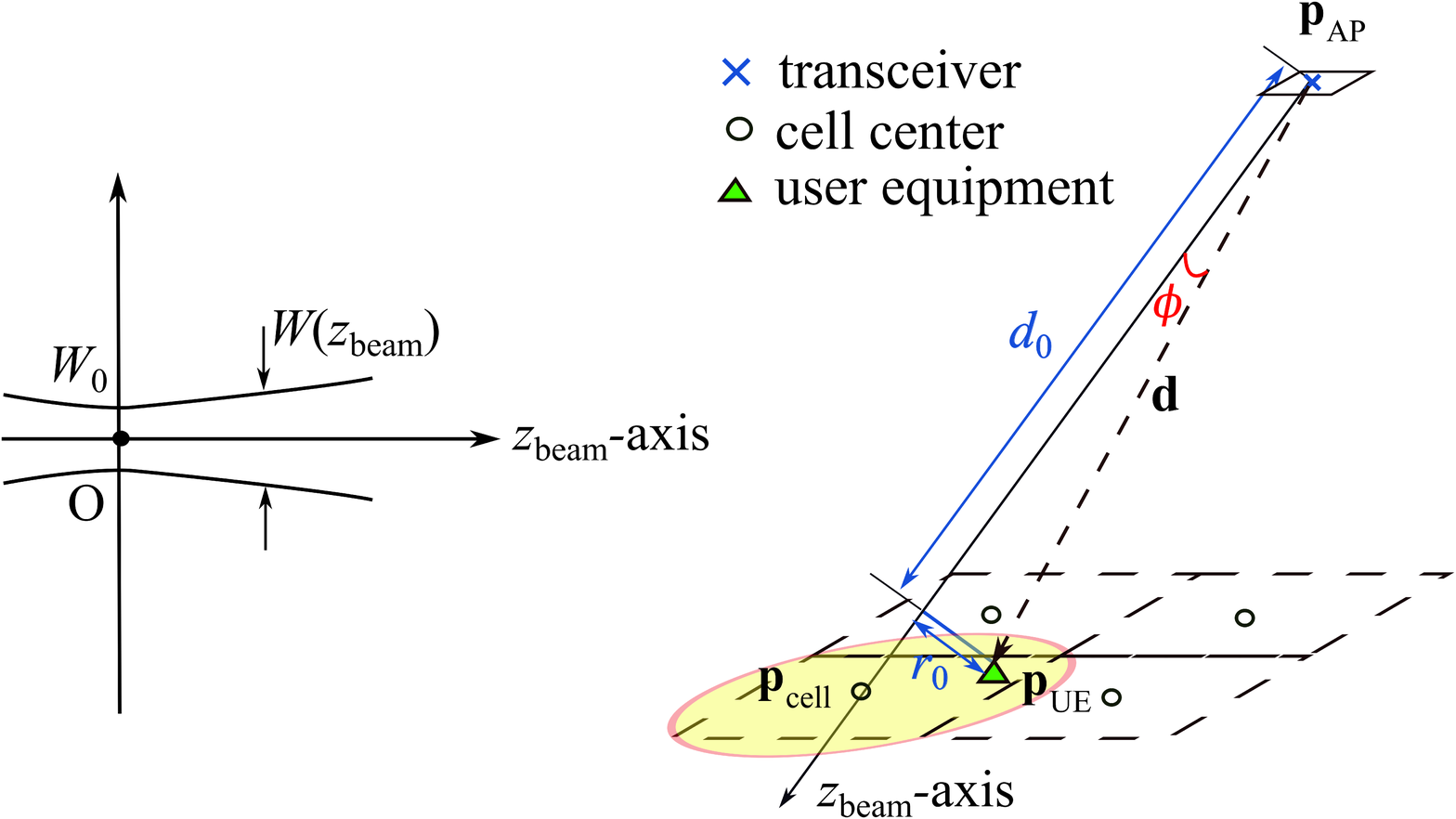}
\caption{Geometrical representation of the elliptical Gaussian beam.}
\label{fig_gaussianbeam}
\vspace{-20pt}
\end{figure}

Depending on the bias current, the VCSEL output beam profile can be Gaussian \cite{Loic_GaussianBeamVCSEL}.
The geometrical representation of the Gaussian beam is shown in Fig. \ref{fig_gaussianbeam} and the Gaussian beam intensity can be expressed as given in \cite{BeamOptics}:
\begin{equation}
    I(r_0,d_0) =\frac{2P_{\rm{tx,opt}}}{\pi W^2(d_0)} \exp{ \Big(-\frac{2r^2_0}{W^2(d_0)}} \Big),
    \label{eq_GBintensity}
\end{equation}
where $P_{\rm{tx,opt}}$ is the transmitted optical power of a single beam and $r_0$ is the distance from the UE to the beam axis, which is represented as $z_{\rm{beam}}$-axis; the distance from the Tx to the UE along the beam axis is given as $d_0$; the beam width at $z_{\rm{beam}}=d_0$ is denoted as $W(d_0)$ and it can be obtained as:
\begin{equation}
    W(d_0)=W_0\sqrt{1+\left(\frac{\lambda d_0}{\pi W^2_0}\right)^2},
\end{equation}
where $\lambda$ is the operating wavelength of the VCSEL. The beam waist is denoted as $W_{0}=\frac{\lambda}{\pi \theta_{\rm{beam}}}$, where $\theta_{\rm{beam}}$ is the  divergence angle. The relation between the beam divergence and the angle for full width at half maximum (FWHM) intensity points, $\theta_{\rm{FWHM}}$, is given as $\theta_{\rm{beam}}=\theta_{\rm{FWHM}}/\sqrt{2\ln(2)}$, where $\ln(\cdot)$ represents the natural logarithm.
The distance vector from the VCSEL to the UE is denoted as $\bf{d}=\bf{p}_{\rm{UE}}-\bf{p}_{\rm{tx}}$ and the distance is $d =\norm{\bf{d}}$. The radiance angle of the Tx is denoted as $\phi=\cos^{-1}\frac{\bf{n}_{\rm{tx}}\cdot\bf{d}}{d}$. It should be noted that $r_0=d\sin{\phi}$ and $d_0=d\cos{\phi}$. Hence, the intensity of the beam at the position of the UE can be reformulated as:
\begin{equation}
    I(d,\phi) =\frac{2P_{\rm{tx,opt}}}{\pi W^2(d\cos\phi)} \exp{ \Big(-\frac{2d^2\sin^2{\phi}}{W^2(d\cos\phi)}} \Big).
\end{equation}
The received optical power at the Rx of a UE is denoted as:
\begin{align}
\begin{split}
P_{\rm{rx,UE}}&=I(d,\phi)A_{\rm{eff}}G_{\rm{APD}}\cos(\psi)\mathrm{rect}\left(\frac{\psi}{\Psi_{\rm c}}\right)\\
&=\frac{2\cos(\psi)P_{\rm{tx,opt}}A_{\rm{eff}}G_{\rm{APD}}}{\pi W^2(d\cos\phi)} \exp \Big(-\frac{2d^2\sin^2{\phi}}{W^2(d\cos\phi)} \Big)\mathrm{rect}\left(\frac{\psi}{\Psi_{\rm c}}\right),
\label{eq_RrxUE}
\end{split}
\end{align}
where $G_{\rm{APD}}$ is the gain of the APD; $A_{\rm{eff}}$ is the effective area of the APD, and $\psi$ is the angle between the normal vector of Rx on the UE, i.e., $\bf{n}_{\rm{UE}}$, and the distance vector ${\bf{d}}$. Therefore, $\psi=\cos^{-1}\frac{\bf{n}_{\rm{UE}}\cdot(-\bf{d})}{d}$. Furthermore, $\mathrm{rect}(\frac{\psi}{\Psi_{\rm c}})=1$, for $0\leq\psi\leq\Psi_{\rm c}$ and $0$ otherwise, where $\Psi_{\rm c}$ is the field of view (FOV) of the receiver. Due to the eye safety consideration, the transmit optical power should be limited. A detailed discussion about eye safety is given in the Appendix, where the beam wavelength $\lambda$ is chosen to be \mbox{1550} nm. The maximum allowable transmitted power for $\theta_{\rm{FWHM}}$ of $2^\circ$, $4^\circ$ and $6^\circ$ are 19, 60 and 129 mW, respectively. 

In this system, the Tx and Rx use intensity modulation (IM) and direct detection (DD), respectively, and the transmit signals are required to be positive and real. The direct current biased optical (DCO)-OFDM is used in this study to achieve high data rates. The sequence number of OFDM subcarriers is denoted by $m \in \{0, \dots, M-1\}$, where $M$ is an even and positive integer which denotes the number of OFDM subcarriers. To ensure real and positive signals, two constraints should be satisfied: i) $X(0) = X(M/2) = 0$, and ii) the Hermitian symmetry constraint, i.e., $X(m) = X^*(M-m)$, for $m\neq0$, where $(\cdot)^*$ denotes the complex conjugate operator  \cite{DPO}. Hence, the effective subcarrier set bearing information data is defined as $\mathcal{M}_{\rm e} = \{m| m \in [1, M/2-1], m \in \mathbb{N}\}$, where $\mathbb{N}$ is the set of natural numbers. For the DCO-OFDM signal, $x_{\rm{DC}}=\kappa\sqrt{P_{\rm{elec}}}$, where $x_{\rm{DC}}$ is the DC bias, $P_{\rm{elec}}$ is the electrical power and $\kappa$ is the conversion factor. By setting $\kappa=3$, it is guaranteed that less than $0.3\%$ of the signals are clipped and therefore the clipping noise is neglectable \cite{MDSTCOM2018}. Therefore, the eletrical SNR of the UE at each effective subcarrier $m$ can be denoted as:
\begin{equation}
\gamma_{m}=
    \frac{(R_{\rm{APD}}P_{\rm{rx,UE}})^2}{(M-2)\kappa^{2}\sigma_{\rm{n}}^2}  , \ \  \ m \in \{1,2, \dots, M/2-1\},
    \label{SNREquation}
\end{equation}
where $P_{\rm{rx,UE}}$ is given in \eqref{eq_RrxUE}; $R_{\rm{APD}}$ is the APD responsivity; $\sigma_{\rm{n}}^2$ is the total noise power for each subcarrier and the detailed discussion of $\sigma_{\rm{n}}^2$ will be given in the next subsection. Based on the Shannon capacity, the data rate for the UE can be expressed as \cite{Dimitrov2013}:
\begin{equation}
\zeta = \sum\limits_{m=1}^{M/2-1}\frac{B_{\rm{L}}}{M} \log_2(1+\gamma_m) =\frac{(M/2-1)}{M}B_{\rm{L}} \log_2\Big(1+\frac{(R_{\rm{APD}}P_{\rm{rx,UE}})^2}{(M-2)\kappa^{2}\sigma_{\rm{n}}^2}\Big)
\label{eq_SE},
\end{equation}
where $B_{\rm{L}}$ is the baseband modulation bandwidth. 
\vspace{-10pt}
\subsection{Receiver Noise}
\begin{table}[!b]
			\centering
			\caption{APD parameters.  } 
			\label{table_APD}
			\vspace{-8pt}
			{\raggedright
				\vspace{4pt} \noindent
				\begin{tabular}{p{130pt}|p{25pt}|p{130pt}}
					\hline
					\parbox{130pt}{\centering{\small Parameter}} & \parbox{25pt}{\centering{\small Symbol}} & \parbox{130pt}{\centering{\small Value}} \\
					\hline
					\hline
					\parbox{130pt}{\raggedright{\small Bandwidth}} & \parbox{25pt}{\centering{\small $B_{\rm{L}}$}} & \parbox{130pt}{\centering{\small $1.5$ GHz}} \\
					\hline
					\parbox{130pt}{\raggedright{\small Spectral response range}} & \parbox{25pt}{\centering{\small -}} & \parbox{130pt}{\centering{\small $950$ to $1700$ nm }} \\
					\hline
					\parbox{130pt}{\raggedright{\small Peak sensitivity wavelength}} & \parbox{25pt}{\centering{\small -}} & \parbox{130pt}{\centering{\small $1550$ nm}} \\
					\hline
					\parbox{130pt}{\raggedright{\small Effective area of APD}} & \parbox{25pt}{\centering{\small $A_{\rm{eff}}$}} & \parbox{130pt}{\centering{\small $\pi \times 0.25 \times 0.25 \times 10^{-4}$ m$^2$}} \\
					\hline
					\parbox{130pt}{\raggedright{\small Receiver FOV}} & \parbox{25pt}{\centering{\small $\Psi_{\rm{c}}$}} & \parbox{130pt}{\centering{\small $60^\circ$ }} \\
					\hline
					\parbox{130pt}{\raggedright{\small Gain of APD}} & \parbox{25pt}{\centering{\small $G_{\rm{APD}}$}} & \parbox{130pt}{\centering{\small $30$ }} \\
					\hline
					\parbox{130pt}{\raggedright{\small Responsivity}} & \parbox{25pt}{\centering{\small $R_{\rm{APD}}$}} & \parbox{130pt}{\centering{\small $0.9$ A/W}} \\
					\hline
					
				\parbox{130pt}{\raggedright{\small Laser noise}} & \parbox{25pt}{\centering{\small RIN}} & \parbox{130pt}{\centering{\small $-155$ dB/Hz }}\\
					\hline
				\end{tabular}
                }
\end{table}
At the receiver, we utilize a high-bandwidth Indium gallium arsenide (InGaAs) APD (G8931-10), where it can work in the spectral range of $950$ nm to $1700$ nm. The peak sensitivity wavelength is $1550$ nm. Parameters of this APD are given in Table~\ref{table_APD} \cite{dataSheet1550}. It is noted that the receiver bandwidth inversely depends on the capacitance of APD, i.e., $B_{\rm{L}}=1/(2\pi R_{\rm{F}}C_{\rm{T}})$, where $C_{\rm{T}}$ is the capacitance of the APD and $R_{\rm{F}}$ is the feedback resistor of the transimpedance amplifier. Therefore, small-area APDs are required to achieve wide bandwidth at the receiver. However, for small-area APDs, a lens is required at the receiver to collect enough power and consequently enhance the SNR. Optical receivers that use an APD are able to provide high SNR. This enhancement in SNR is due to the internal gain of the APD, $G_{\rm{APD}}$. If the noise of the receiver was independent of the internal gain, the SNR would increase by a factor of $G_{\rm{APD}}^2$. Unfortunately, the noise depends on $G_{\rm{APD}}$ so that the SNR improvement is less that $G_{\rm{APD}}^2$ \cite{agrawal2005lightwave}. The noise at the receiver is due to three factors, namely thermal noise, shot noise and relative intensity noise (RIN). In the following, we will characterize the different types of noise at the receiver. 

\subsubsection{Thermal Noise}
Thermal noise is characterized for an APD the same as other PDs. 
The power spectral density (PSD) of thermal noise is given as \cite{WirelessInfrared}:
\begin{equation}
    S_{\rm{thermal}}(f)=\frac{4k_{\rm{B}}T}{R_{\rm{F}}},
\end{equation}
where $k_{\rm{B}}$ is the Boltzmann's constant, $T$ is absolute temperature.

\subsubsection{Shot Noise}
In optical communication, shot noise is due to the random nature of photon arrivals with an average rate determined by the incidence optical power.  
The PSD of the shot noise for an APD is given as:
\begin{equation}
\label{ShotNoise}
    S_{\rm{shot}}(f)=2qG_{\rm{APD}}^2F_{\rm{A}}R_{\rm{APD}}(P_{\rm{rx,UE}}+P_{\rm{n}}),
\end{equation}
where $q$ denotes the electron charge and the average ambient power is represented by $P_{\rm{n}}$. 
We note that under the condition of $P_{\rm{n}}\gg P_{\rm{rx,UE}}$ the shot noise is signal independent and is only affected by the ambient light. In \eqref{ShotNoise}, $F_{\rm{A}}$ is called the excess noise factor and can be obtained as:
\begin{equation}
    F_{\rm{A}}=k_{\rm{A}}G_{\rm{APD}}+(1-k_{\rm{A}})\left(2-\frac{1}{G_{\rm{APD}}}\right),
\end{equation}
where $0<k_{\rm{A}}<1$ is a dimensionless parameter. 

\subsubsection{Relative Intensity Noise}
RIN is another type of noise induced mainly due to the instability in the transmit power of the VCSEL. Cavity variation and fluctuations in the laser gain are two major contributors to the RIN. The variance of power fluctuations is given as \cite{5466150}:
\begin{equation}
\label{RINnoise}
    \sigma_{\rm{I}}^2=(R_{\rm{APD}}P_{\rm{rx,UE}}r_{\rm{I}})^2,
\end{equation}
where $r_{\rm{I}}$, is a measure of the noise level of the optical signal and is given as:
\begin{equation}
    r_{\rm{I}}^2=\int_{-\infty}^{\infty}{\rm{RIN}}(f) {\rm{d}}f,
\end{equation}
where ${\rm{RIN}}(f)$ is the intensity noise spectrum. For a limited bandwidth receiver, the above integral should be calculated over the receiver bandwidth. It is noted that $r_{\rm{I}}$ is simply the inverse of the transmit optical power. 
For simplicity, we assume that ${\rm{RIN}}(f)$ is a constant value over the whole bandwidth, i.e., ${\rm{RIN}}(f)={\rm{RIN}}$. Hence, the PSD of the relative noise intensity can be denoted as:
\begin{equation}
   S_{\rm{RIN}}(f)={\rm{RIN}}(R_{\rm{APD}}P_{\rm{rx,UE}})^2.
\end{equation}
\subsubsection{Total Noise}
Finally, the total noise power for each subcarrier at the receiver can be expressed as:
\begin{equation}
    \sigma_{\rm{n}}^2=\frac{4k_{\rm{B}}T}{R_{\rm{F}}}B_{\rm{L}}/M+2qG_{\rm{APD}}^2F_{\rm{A}}R_{\rm{APD}}(P_{\rm{rx,UE}}+P_{\rm{n}})B_{\rm{L}}/M+{{\rm{RIN}}(R_{\rm{APD}}P_{\rm{rx,UE}})^2B_{\rm{L}}/M}.
\end{equation}

\subsection{Retroreflector}
\label{SectionRetroreflector}
A retroreflector can reflect part of the incident beam back to the direction it came from. The retroreflector can be realized by a CCR, retroreflective sheeting or even retroreflective spray paint \cite{RetroAlex}. In this paper, the property of a CCR is studied. As the distance between the Tx and the CCR, $d$, is much larger than the size of a CCR, all the incident rays can be 
\begin{figure*}[!ht]
	\centering
	\begin{subfigure}[!ht]{0.4\textwidth}
		\centering
		\includegraphics[height=0.5\columnwidth]{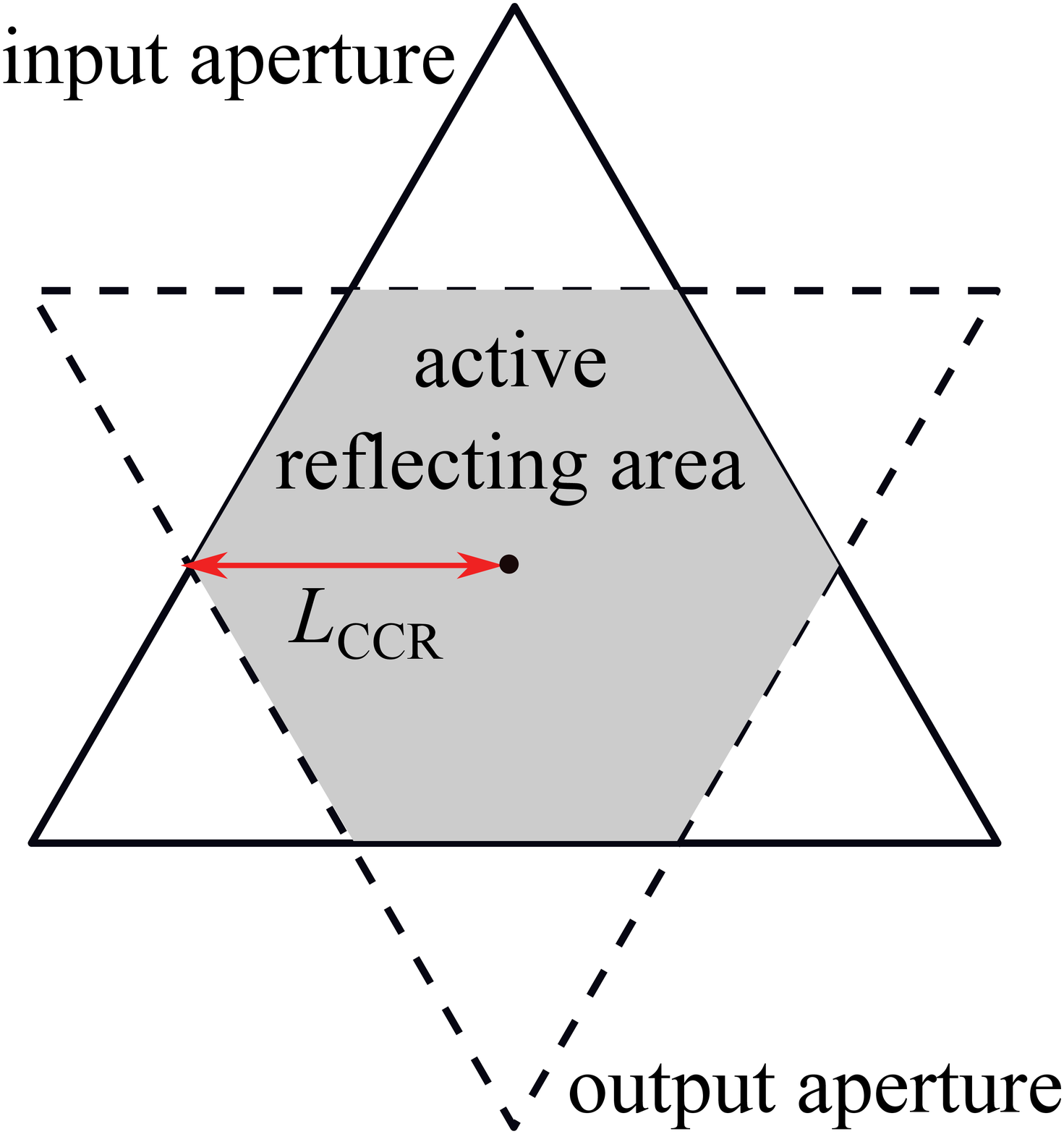}
		\caption{X-Y view}
		\label{fig_active_reflecting_area}
	\end{subfigure}%
	~ 
	\begin{subfigure}[!ht]{0.4\textwidth}
		\centering
		\includegraphics[height=0.5\columnwidth]{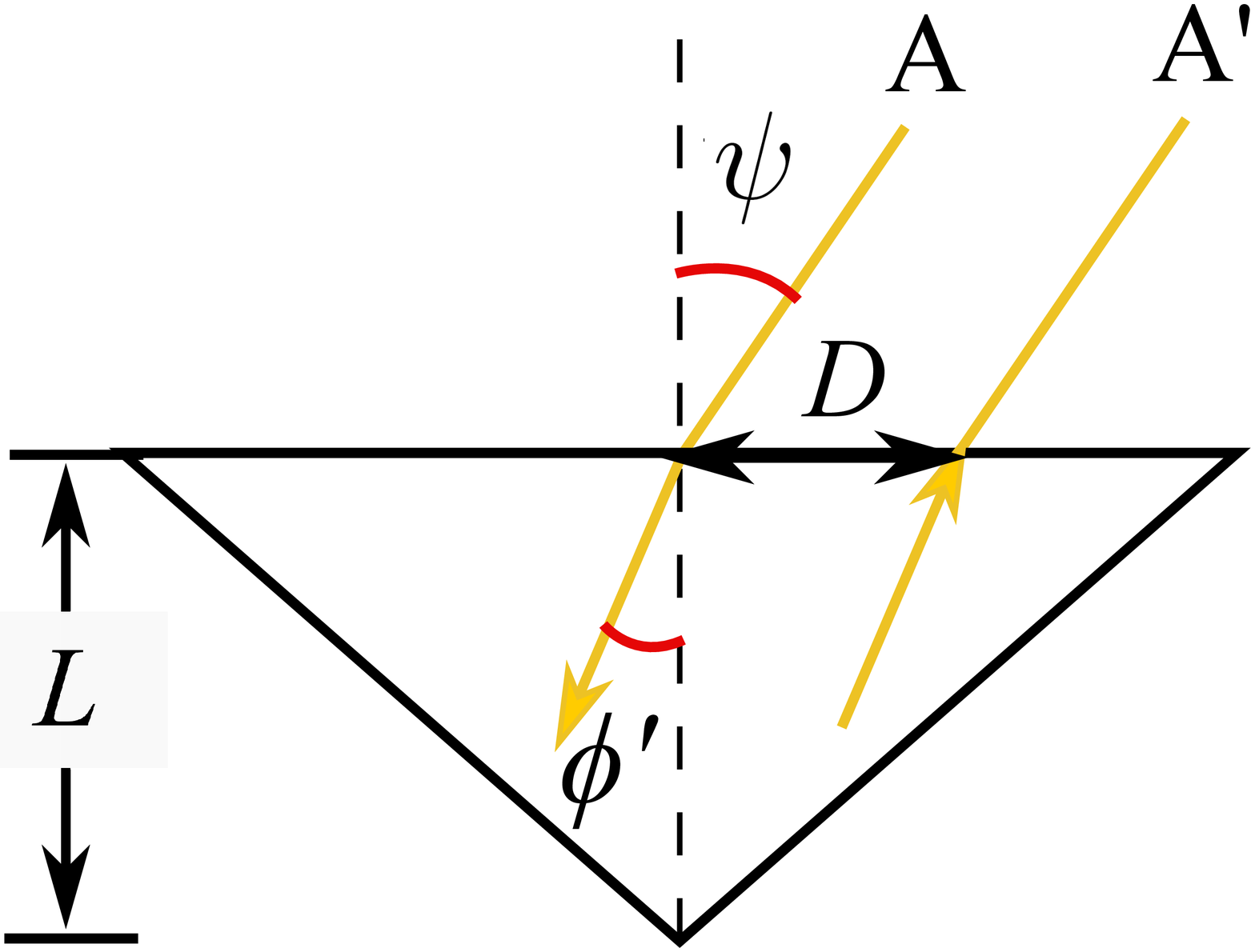}
		\caption{X-Z view}
		\label{fig_retro_xzView}
	\end{subfigure}
	\caption{Corner-Cube retroreflector }
	\vspace{-10pt}
\end{figure*}
considered to be parallel with an incident angle of $\psi$. 
According to Snell's law, the refracted angle $\phi'=\sin^{-1}\frac{\sin\psi}{n_{\rm{re}}}$, where $n_{\rm{re}}$ is the refractive index of the retroreflector. Fig. \ref{fig_active_reflecting_area} shows the result for a triangular corner-cube reflector at normal incidence. The solid line is the input aperture, which is the shape of the CCR face, and the dotted line is the output aperture which is the outline of the retroreflected beam. The overlap area of the input and output aperture is the active reflecting area, and the incident rays outside this area will not be retroreflected. Fig. \ref{fig_retro_xzView} shows the result for a triangular corner-cube reflector at oblique incidence. 
The displacement of input and output apertures, $D$, can be obtained as:
\begin{equation}
D = 2L\tan\phi',
\end{equation}
where $L$ is the length of the CCR. The size of the active reflecting area mainly depends on $L_{\rm{CCR}}$, which is shown in Fig. \ref{fig_active_reflecting_area}.

\section{Beam Activation}
\label{SectionBeamActivation}
For a VCSEL array system, it is important to determine which beam should be activated to optimize the system performance. The beam selection strategy considered in this study is the signal strength strategy (SSS), which only selects the beam providing the highest received power. Other beam selection strategies can be applied to activate two or more beams simultaneously to achieve better performance. However, two or more beams consume more power, which may not be energy efficient. The trade of between spectrum and energy efficiency will be considered in our future work. Therefore, the index of the serving beam for the user is expressed as:
\begin{equation}
    I = \argmax_{n \in \mathcal{N}} P_{\rm{rx,UE}}^{n},
    \label{beam_index1}
\end{equation}
where $\mathcal{N}$ is the set of beams and $P_{\rm{rx,UE}}^{n}$ is the received optical power of the user from the $n$-th beam. To solve \eqref{beam_index1}, the knowledge of user location is required. The cell size is very small in the VCSEL array system, therefore, an accurate localization technique is needed. For the mobile UE, the user tracking system is required to update the location of the UE quickly and frequently. The VLP can easily achieve sub-meter accuracy. However, most VLP localization technologies utilize the downlink transmission of the VLC system {\cite{Koonen_JLT_2018}}, which requires extra light setting for the proposed system. In addition, the location information needs to be processed and sent back to the server, which causes extra delays to the real-time beam activation system. Two beam activation schemes are proposed in this study to address the power consumption and delay issue. The first method is a passive beam activation scheme, which uses a CCR to obtain a power matrix and then find the serving beam. The second method uses the ODTx in the uplink communication. The details of these two schemes are presented in the following subsections. 

\subsection{Systems with a CCR}
\begin{figure*}[!t]
 \centering
 \includegraphics[width=0.5\textwidth]{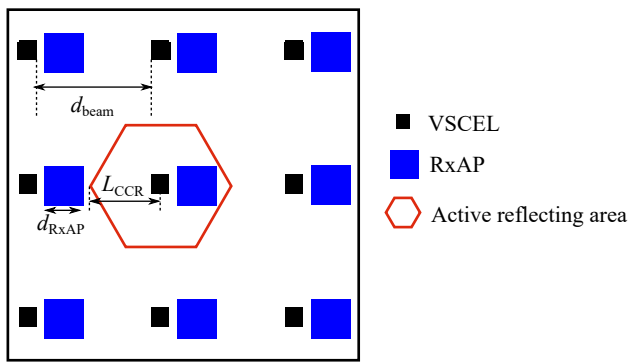}
\caption{ The setup of the AP in the VCSEL array system using a CCR $(N_{\rm{beam}}=9)$. }
\label{fig_APsetup}
\vspace{-20pt}
\end{figure*}

A CCR can reflect part of the incident light to the opposite direction. Based on this property, a passive beam activation is proposed to reduce the power consumption of the battery-operated UE and the latency. A CCR will be mounted next to the Rx of a UE and will reflect part of the transmitting light back to the source beam. As shown in Fig. \ref{fig_APsetup}, a PD, which is referred to as RxAP, is mounted next to each source beam to detect the light reflected by the CCR. The area bounded by the red line in Fig. \ref{fig_APsetup} is the active reflecting area, which is the area hit by the retroreflected light. As discussed in Section \ref{SectionRetroreflector}, the size of the active reflecting area is determined by $L_{\rm{CCR}}$, which mainly depends on the size of the retroreflector.
To make sure the light transmitted by the source beam, after being reflected by the CCR, will only be received by the RxAP next to this source beam, the source beams need to be separated from each other with a distance of $d_{\rm{beam}} \ge L_{\rm{CCR}} + d_{\rm{RxAP}}$. In this study, the orientation of the Rx and CCR are assumed to be fixed and vertically upward. However, it should be noted that due to the limitation of the maximum incident angle of the CCR and the rotation of the UE, a single Rx may not always receive the signal and a single CCR may not always reflect the light back to the reversed direction. As a result, when a UE's orientation is considered, multiple Rxs and CCRs should be mounted in different directions to solve the issue. The structure of multiple Rxs and CCRs, which ensures that the system is robust against random orientation, will be a subject for future studies. In terms of beam activation mechanism, before transmitting the data packet, all the beams will send test signals simultaneously. By monitoring the power matrix, which is the power received by the array of RxAPs, the beam corresponds to the RxAP receiving the maximum reflected power will be activated to transmit the data packet. Therefore, the index of the source beam for the UE in \eqref{beam_index1} can be reformulated as:
\begin{equation}
    I = \argmax_{n \in \mathcal{N} } P_{\rm{RxAP}}^{n},
    \label{beam_index2}
\end{equation}
where $P_{\rm{RxAP}}^{n}$ is the received power, reflected by the UE's CCR, of the $n$-th RxAP. 

By covering the front face of the CCR with a liquid crystal display (LCD), the retroreflected light can be modulated \cite{ShaoRetro2018}. In addition, the power consumption of the LCD with its driver circuit is only in the order of tens of $\mu$W. The use of the LCD serves multiple purposes. First, it can be used to determine whether a UE is active and requires data communication. Second, it helps to distinguish the active UE from mirrors or CCRs in the environment. Finally, for a multi-user scenario, the active UEs can be distinguished among each other through modulating the retroreflected light differently.

\begin{figure*}[!t]
    \centering
    \includegraphics[width=0.8\textwidth]{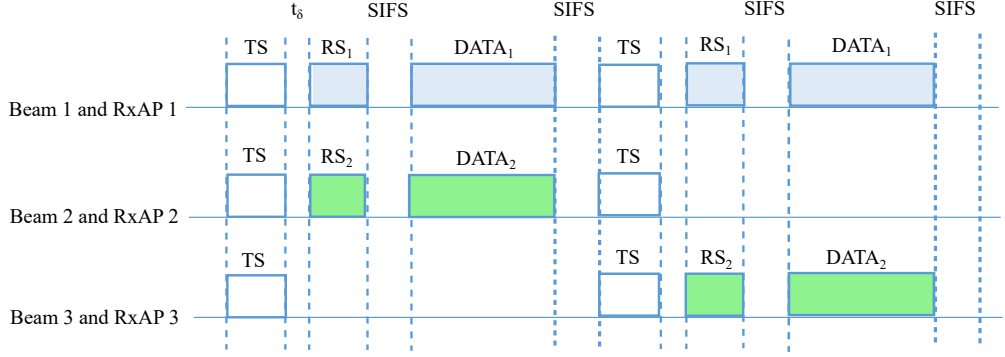}
    \caption{Beam activation mechanism.}
    \label{fig_ActivationMechanism}
    \vspace{-20pt}
\end{figure*}
The beam activation mechanism is shown in Fig. \ref{fig_ActivationMechanism}. At the beginning, all the beams send test signals (TS) simultaneously. It is noted that the TS can be unmodulated light. The UE will retroreflect and modulate the TS into the reflected signal (RS), and the RS will be received by the corresponding RxAP  after a mean propagation delay of $t_\delta$. The beam will be activated based on \eqref{beam_index2}. Then, the data frame will be transmitted by the serving beam after a short inter-frame space (SIFS). In Fig. \ref{fig_ActivationMechanism}, firstly, beams 1, 2, and 3 send TS simultaneously. Then, the RxAP 1 and 2 receive RS retroreflected by the UE 1 and UE 2, respectively, while the RxAP 3 does not receive any reflected signal. Therefore, beams 1 and 2 are activated to transmit data frames to UE 1 and UE 2,  respectively. After the SIFS, all the beams send TS again. Here, the RxAP 3 receives RS retroreflected by the UE 2, which means that UE 2 moved to the cell served by the beam 3. Hence, the beam 3 is activated to transmit data to the UE 2 while the beam 2 is deactivated. As a consequence, real-time tracking of UEs and real-time beam activation can be executed with almost zero-delay for a UE with ultra-low power and no computational capability. In addition, the proposed system with CCR does not first require the estimation of the position of UEs.

\subsubsection*{Signaling Cost Analysis}
The effective throughput can be obtained based on the ratio of the time allocated for data transmission and the total time including that used for the signaling. Accordingly, we have:
\begin{equation}
\begin{aligned}
    &t_{\rm{Data}}=\frac{L_{\rm{Data}}}{\zeta_{\rm{down}}}, \ \ t_{\rm{tot}}=t_{\rm{Data}}+t_{\rm{delay}}, \\
    & t_{\rm{delay}}=t_{\rm{TS}}+t_\delta+t_{\rm{RS}}+
    t_{\rm{SIFS}}+t_{\rm{SIFS}},
\end{aligned}
\label{eq_time_delay}
\end{equation}
where $t_{\rm{Data}}$ is the average data transmission time; $L_{\rm{Data}}$ is the average data packet length; $\zeta_{\rm{down}}$ is the average downlink data rate which can be obtained based on \eqref{eq_SE};  the $t_{\rm{delay}}$ is the delay time; and $t_{\rm{tot}}$ is the total time required to complete the transmission of a packet; $t_{\rm{TS}}$ and $t_{\rm{RS}}$ are the signaling periods; $t_{\rm{SIFS}}$ is the time required for the SIFS. Assuming there is no collision, the effective throughput can be obtained as:
\begin{equation}
    \mathcal{T}_{\rm{eff}}=\frac{t_{\rm{Data}}}{t_{\rm{tot}}}\zeta_{\rm{down}}.
    \label{eq_throughput_CCR}
    \vspace{-20pt}
\end{equation}

\begin{table}[t]
			\centering
			\caption{Parameters of beam activation mechanism.}
			\label{table_BeamActivaiton}
			\vspace{-8pt}
			{\raggedright
				\vspace{4pt} \noindent
				\begin{tabular}{p{150pt}|p{30pt}|p{100pt}}
					\hline
					\parbox{150pt}{\centering{\small Parameter}} & \parbox{30pt}{\centering{\small Symbol}} & \parbox{100pt}{\centering{\small Value}} \\
					\hline
					\hline
					\parbox{150pt}{\raggedright{\small Test signal time}} & \parbox{30pt}{\centering{\small $t_{\rm{TS}}$}} & \parbox{100pt}{\centering{\small $0.3$ micro-seconds}} \\
					\hline
					\parbox{150pt}{\raggedright{\small Reflected signal time}} & \parbox{30pt}{\centering{\small $t_{\rm{RS}}$}} & \parbox{100pt}{\centering{\small $0.3$ micro-seconds}} \\
					\hline
					\parbox{150pt}{\raggedright{\small Average length of data packet \cite{80211axTutorial}}} & \parbox{30pt}{\centering{\small $L_{\rm{Data}}$}} & \parbox{100pt}{\centering{\small {$64$ Kbytes}}} \\
					\hline
					\parbox{150pt}{\raggedright{\small Mean propagation delay \cite{Higgins2009JLT}}} & \parbox{30pt}{\centering{\small $t_\delta$}} & \parbox{100pt}{\centering{\small $3$ ns}} \\
					\hline
					\parbox{150pt}{\raggedright{\small SIFS \cite{80211axTutorial}}} & \parbox{30pt}{\centering{\small SIFS}} & \parbox{100pt}{\centering{\small $2$ micro-seconds}} \\
					\hline
				\end{tabular}
                }
\vspace{-0.5cm}
\end{table}

\subsection{Systems with ODTx on the UE}
\begin{figure*}[!ht]
    \centering
    \includegraphics[width=0.2\textwidth]{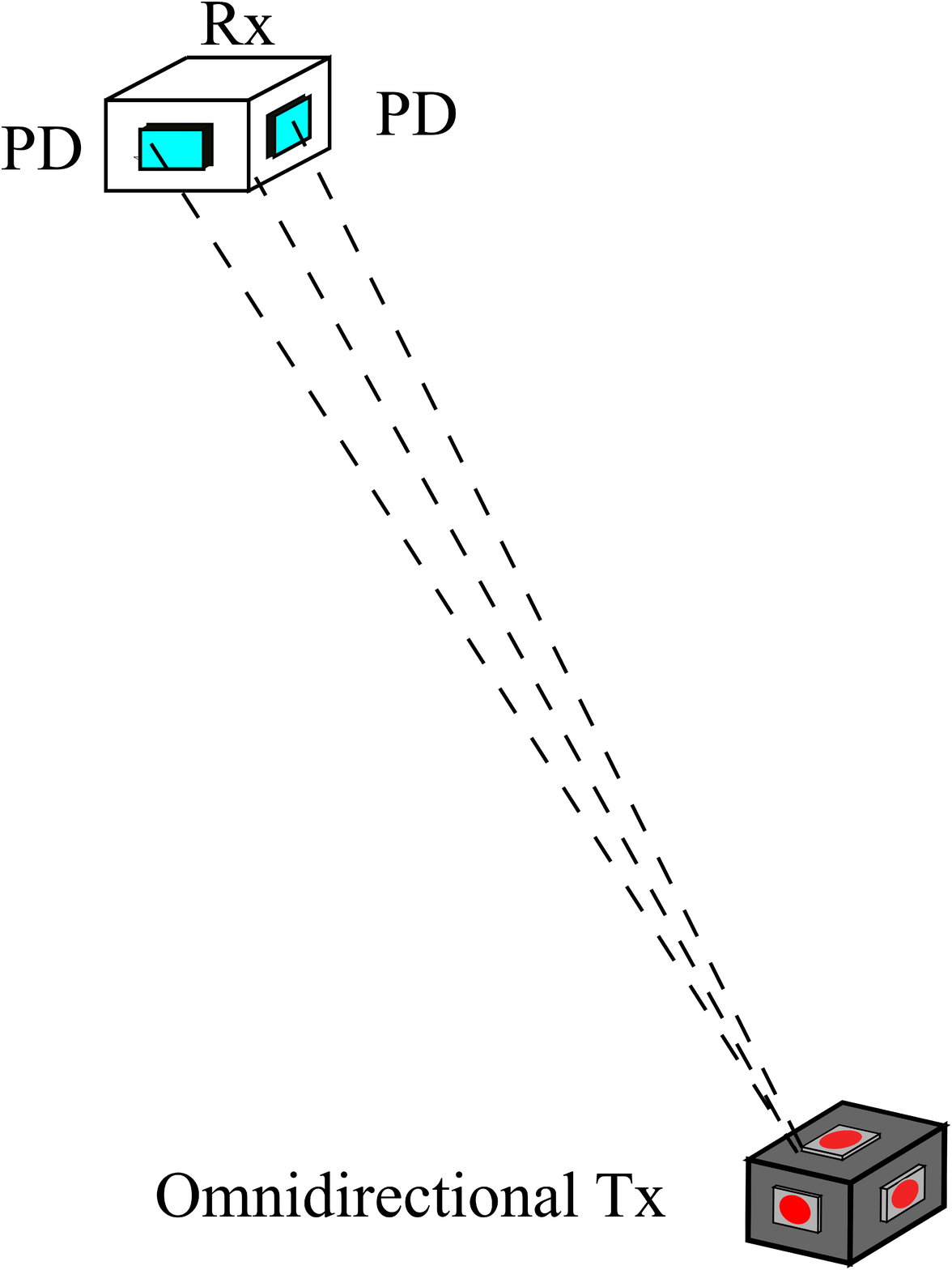}
    \caption{Localization using the ODTx.}
    \label{fig_omniTx}
        \vspace{-10pt}
\end{figure*}


In this study, an alternative scheme is proposed for beam activation by using ODTx for the wireless infrared uplink transmission. An ODTx, which achieves omnidirectional  transmission characteristics by using multiple transmitting elements, is first proposed for LiFi systems in \cite{chen2018infrared}. The structure of this ODTx is shown in Fig. \ref{fig_omniTx}. There is one infrared LED located on each side of the UE. It is assumed that the infrared LED follows Lambertian radiation and $m$ is the Lambertian order which is given as \mbox{$m=2$}. The received optical power of a PD from the ODTx can be written as \cite{chen2018infrared}:
\begin{align}
\centering
	P_{\rm{rx,OD}} =
	\begin{cases}
\frac{(m+1)P_{\rm{tx,OD}}A_{\rm{OD}}n^2_{\rm{ref}}\cos(\psi_{\rm{OD}})}{2{\pi}d^2\sin^2(\Psi_{\rm{OD}})},& \psi_{\rm{OD}}\leq\Psi_{\rm{OD}}\\
0, & \text{otherwise}
	\label{eq_PrxOD}
	\end{cases},
\end{align}
where $R_{\rm{PD}}$ is the PD responsivity and $P_{\rm{tx,OD}}$ is the transmitted optical power of a single infrared LED in the ODTx; $A_{\rm{OD}}$ represents the physical area of the PD; $n_{\rm{ref}}$ denotes the internal refractive index; the incidence angle of the PD is $\psi_{\rm{OD}}$ and the FOV of the PD is $\Psi_{\rm{OD}}$. Based on \eqref{eq_PrxOD}, it can be observed that the received optical power is independent of the irradiance angle of the ODTx, which means that the rotation of the UE will not change $P_{\rm{rx,OD}}$.  Consequently, the received optical power only depends on the position of the UE, $\bf{p}_{\rm{UE}}$. 
 
\begin{figure*}[!b] 
	\centering
	\begin{subfigure}[!ht]{0.45\textwidth}
		\centering
		\includegraphics[height=0.7\columnwidth]{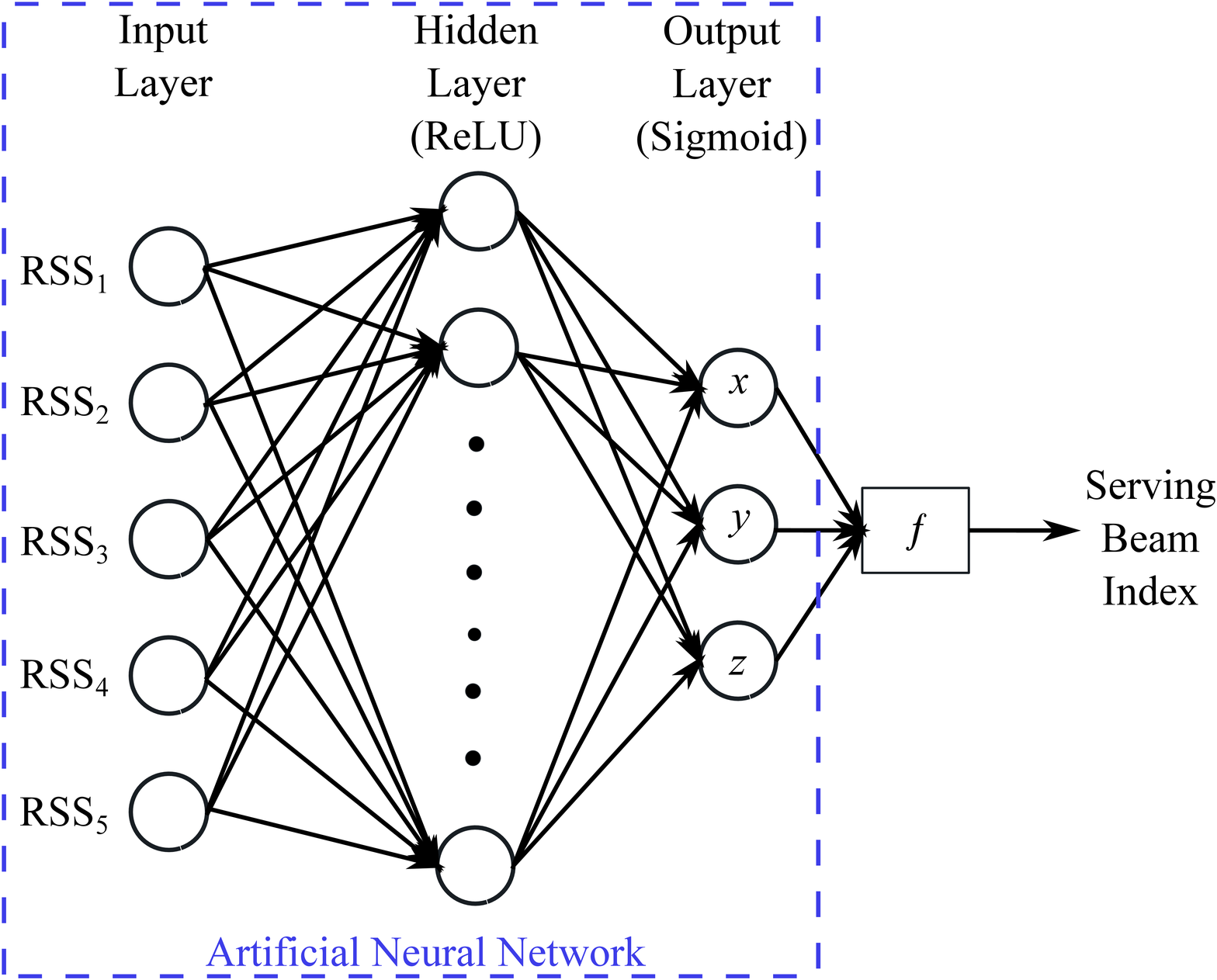}
		\caption{ANN for UE localization}
		\label{fig_NN1}
	\end{subfigure}%
	~ 
	\begin{subfigure}[!ht]{0.45\textwidth}
		\centering
		\includegraphics[height=0.7\columnwidth]{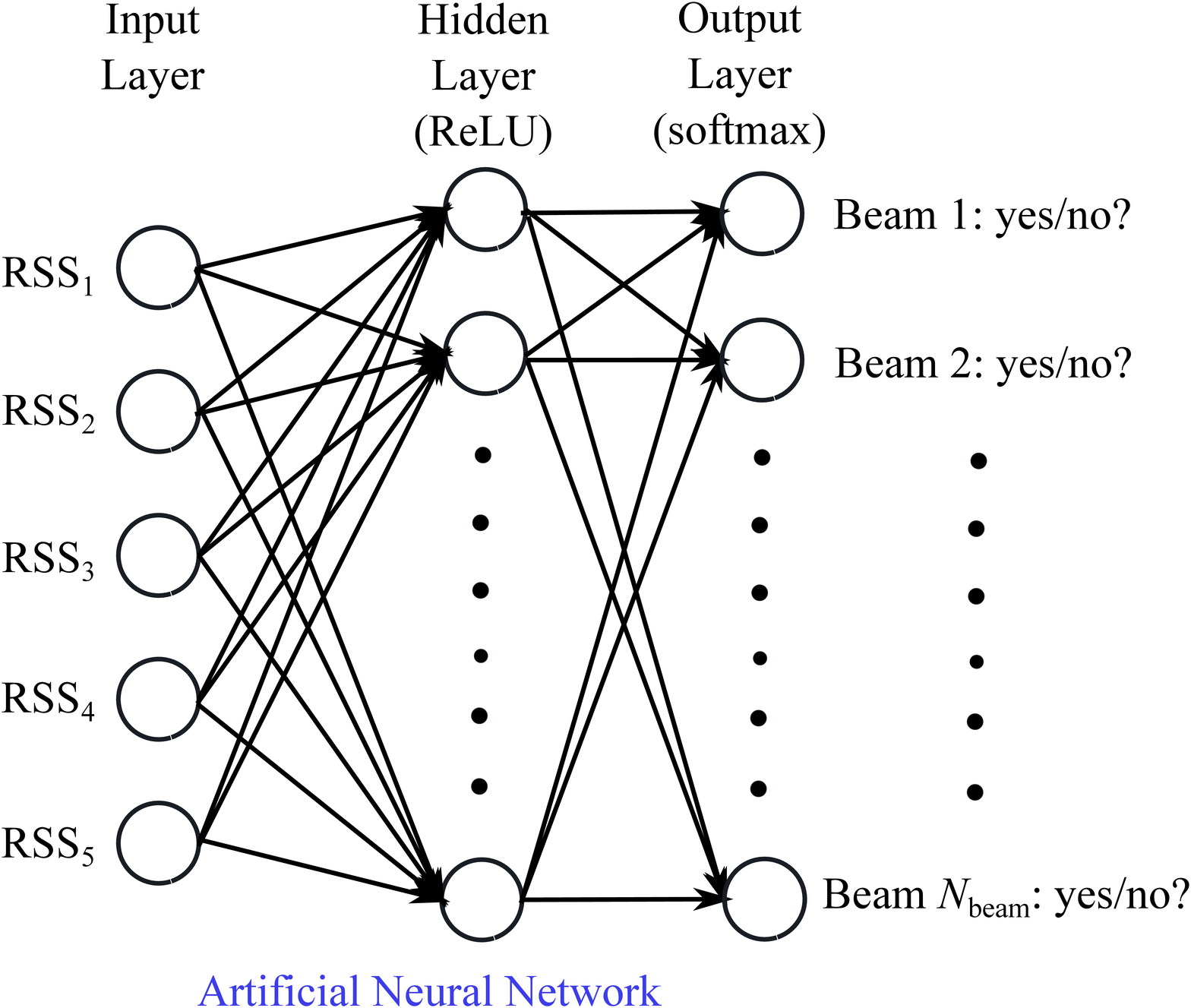}
    	\caption{ANN for beam activation}
    	\label{fig_NN2}
	\end{subfigure}
	\caption{The ANN structures}
	\label{fig_NN}
	\vspace{-20pt}
\end{figure*}

By mounting multiple PDs in different orientations on the ceiling, multiple RSS values, $P_{\rm{rx,OD}}$, can be collected. In indoor positioning systems, RSS-based methods have been widely adopted since RSS values are easy to obtain \cite{ZhuangSurveyPositioning}. In this study, the ANN algorithm is used to estimate the user position. The main motivation of using ANN is that it can estimate the beam activation directly without estimating the UE location, which reduces the latency. Other methods require the prediction of the user position using techniques, such as fingerprinting, proximity and trilateration first and then complete the beam activation.  Conventionally, the ANN is adopted to estimate the user position and after finding the position of the UE, $\bf{p}_{\rm{UE}}$, the beam that needs to be activated can be obtained based on \eqref{beam_index1}. Fig. \ref{fig_NN1} shows the layout of the ANN for the UE localization. The input of the ANN is the RSS from 5 receiving PDs facing in different directions. There are $N_{\rm{hidden}}$ neurons in the hidden layer and the activation function is a rectified linear unit (ReLU). In the output layer, the sigmoid activation function is adopted and the output is the position of the UE $\bf{p}_{\rm{UE}}$, which should be normalized. Then, the index of the serving beam can be calculated based on \eqref{beam_index1}.  In this study, the proposed ANN predicts the index of serving beam directly without the need of positioning first. The layout of the ANN for beam activation is shown in Fig. \ref{fig_NN2}.  The input and hidden layer is the same as the counterpart in the ANN for UE localization, while the output layer uses the softmax activation function and the number of output neurons equals the number of beams $N_{\rm{beam}}$. The value of each output neuron represents the probability of activating the corresponding beam. As the SSS is adopted here, the beam with highest probability should be selected as the serving beam for the UE. The beam activation can be executed in two steps: the offline training of ANN and then the online activating of beams. The RSS data in this study can be generated based on the ideal channel model in \eqref{eq_PrxOD}. However, the Lambertian irradiation model is not exactly accurate in real applications. To ensure the accuracy of the beam activation, the ANN can be pre-trained using the data from the ideal model and then trained with the on-site data \cite{ZhangVLPANN}.

Users can access the uplink and receive the data on downlink simultaneously since the uplink and downlink are operating on different wavelengths and there would be no interference between them. It is assumed that the uplink is always active for data or TS transmission. By using the collected uplink RSS values, the index of the serving beam can be predicted.  However, the delay time, $t_{\rm{delay}}$, of the system is caused by the signal propagation from the ODTx to the AP and the ANN processing time to estimate the active beam index. The signal propagation time, which is in the order of tens of ns, can be ignored when compared with the ANN processing time. Hence, the $t_{\rm{delay}}$ is mainly determined by the ANN processing. For a static user, the effect of $t_{\rm{delay}}$ can be ignored. If the user is moving, $t_{\rm{delay}}$ should be taken into account when evaluating the system performance. It is noted that this method can be still applied to mobile users with low speed however for those users with high speed the delay due to ANN mechanism can cause the location information be outdated and a wrong beam might be activated. When there are multiple UEs, multiple access schemes such as time division multiple access (TDMA), frequency division multiple access (FDMA), or carrier sense multiple access with collision avoidance (CSMA/CA) should be adopted.

\vspace{-15pt}
\section{System Level Analysis}
\label{SectionSystemLevelAnalysis}
\vspace{-5pt}
\subsection{System with single user}
In a VCSEL array system, the total number of beams is denoted as $N_{\rm{beam}}$ and the beams are pointing in different directions. Therefore, the channel gain distribution is different for each beam. The central beam, which points vertically downward, provides the best channel as it has the shortest distance to the UE.  Hence, initially, we will start with the central beam and then we extend the analysis to the whole system. In a single user system, the SNR and average data rate of the $n$-th beam are represented as $\gamma^n_{\rm{single}}$ and $\Bar{\zeta}^n_{\rm{single}}$, respectively. The central beam is assumed to be the $1^{\rm{st}}$ beam and its SNR and average data rate are denoted as $\gamma^1_{\rm{single}}$ and $\Bar{\zeta}^1_{\rm{single}}$, respectively. For the central beam, the horizontal distance between an active user and the cell center is denoted as $r$. The PDF of $r$ can therefore be given as:
\vspace{-5pt}
\begin{equation}
    f_{r}(r)=\frac{2r}{R^2}, \ \ \ r \leq R,
    \vspace{-5pt}
\end{equation}
where $R$ is the radius of the service region of the central beam and $R=\sqrt{d^2_{\rm{cell}}/\pi}$. As the central beam is pointing vertically downward,  \mbox{$\cos\phi=\cos\psi=h/\sqrt{h^2+r^2}$}, $\sin\phi=r/\sqrt{h^2+r^2}$ and $d=\sqrt{h^2+r^2}$. Hence, for UEs served by the central beam, \eqref{eq_RrxUE} can be reformulated as:
\vspace{-5pt}
\begin{equation}
P_{\rm{rx,UE}}(r)=\frac{2P_{\rm{tx,opt}}A_{\rm{eff}}G_{\rm{APD}}}{ \pi W^2(h)}\frac{h}{\sqrt{h^2+r^2}} \exp{\Big( -\frac{2r^2}{W^2(h)}}\Big) .
\vspace{-5pt}
\end{equation}
For a single user case, the system is noise-limited and the there is no interference. Therefore, the SNR given in \eqref{SNREquation} for the user served by the central beam can be obtained as:
\vspace{-5pt}
\begin{equation}
    \gamma^1_{\rm{single}}(r)
    =\frac{\big(R_{\rm{APD}}P_{\rm{rx,UE}}(r)\big)^2}{(M-2)\kappa^{2}\sigma^2_{\rm{n}}}
    =\frac{\gamma_0}{h^2+r^2} \exp{\big( -\frac{4r^2}{W^2(h)}}\big).
    \label{eq_snr_central_cell}
\vspace{-5pt}
\end{equation}
where $\gamma_0=\big( \frac{2R_{\rm{APD}} P_{\rm{tx,opt}}A_{\rm{eff}}G_{\rm{APD}}h}{ \pi W^2(h)\sqrt{M-2}\kappa \sigma_{\rm{n}}} \big)^2$. The SNR in decibel (dB) can be expressed as:
\vspace{-5pt}
\begin{equation}
\gamma^1_{\rm{db}}(r)=10\log_{10}\big(\gamma^1_{\rm{single}}(r)\big).
\label{eq_snr1_db}
\vspace{-10pt}
\end{equation} 
Recalling from the fundamental theorem of a function of a random variable \cite{Papoulis}, the PDF of $\gamma^1_{\rm{db}}$ can be written as:
\vspace{-10pt}
\begin{equation}
    f_{\gamma}(\gamma)
    =\frac{f_r(r_0)}{|\frac{{\rm{d}}}{{\rm{d}}r}\gamma^1_{\rm{db}}(r)|_{r=r_0}}
    = \frac{\log(10)(h^2+r_0^2)W^2(h)}{10R^2\big(4h^2+4r_0^2+W^2(h)\big)},
    \label{eq_pdfsnr}
    \vspace{-10pt}
\end{equation}
where 
\begin{equation}
 \resizebox{0.9\hsize}{!}{$
    r_0 
    =f_r^{-1}(\gamma) 
    =\frac{1}{2}\sqrt{W^2(h)W_k\left(\frac{4\gamma_0\exp\Big(4h^2/W^2(h)-y\log(10)/10\Big)}{W^2(h)}\right)-4h^2},
     \ \ \ \gamma^1_{\rm{db}}(R) \leq \gamma \leq \gamma^1_{\rm{db}}(0) ,
    $}
    \vspace{-5pt}
\end{equation}
where $W_k(\cdot)$ is the Lambert W function, also called as the omega function or product logarithm. As $0\leq r_0 \leq R \ll h$, the PDF of $\gamma^1_{\rm{db}}$ in \eqref{eq_pdfsnr} can be approximated as:
\begin{equation}
    f_{\gamma}(\gamma)\approx \frac{\log(10)(h^2)W^2(h)}{10R^2\big(4h^2+W^2(h)\big)},
         \ \ \ \gamma^1_{\rm{db}}(R) \leq \gamma \leq \gamma^1_{\rm{db}}(0) ,
    \label{eq_pdfsnr_approx}
\end{equation}
Therefore, the approximated PDF of $\gamma^1_{\rm{db}}$ is a uniform distribution function between $\gamma^1_{\rm{db}}(R)$ and $\gamma^1_{\rm{db}}(0)$,  where  $\gamma^1_{\rm{db}}(R)$ and $\gamma^1_{\rm{db}}(0)$ can be obtained from \eqref{eq_snr1_db}.

Based on \eqref{eq_SE} and \eqref{eq_snr_central_cell}, the data rate of the central beam at radius $r$ can be represented as:
\begin{equation}
    \zeta^1_{\rm{single}}(r)=\frac{M/2-1}{M}B_{\rm{L}} \log_2\Big(1+\frac{\gamma_0}{h^2+r^2} \exp{\big( -\frac{4r^2}{W^2(h)}}\big)\Big).
\end{equation}
Therefore, the average data rate of the central beam can be written as:
\begin{equation}
    \begin{split}
        \Bar{\zeta}_{\rm{single}}^1 &= \mathbb{E} \big[ \zeta_{\rm{single}}^1(r) \big] \\
         &= \int_0^R \frac{M/2-1}{M}B_{\rm{L}} \log_2\Big(1+\frac{\gamma_0}{h^2+r^2} \exp{\big( -\frac{4r^2}{W^2(h)}}\big)\Big) f_{r}(r) \ {\rm{d}}r\\
         &=\frac{M-2}{2M}B_{\rm{L}} \int_0^R \log_2\Big(1+\frac{\gamma_0}{h^2+r^2} \exp{\big( -\frac{4r^2}{W^2(h)}}\big)\Big)\frac{2r}{R^2} \ {\rm{d}}r\\
         & \approx	\frac{M-2}{2\log(2)MR^2}B_{\rm{L}} \int_0^R 2r\log\Big( \frac{\gamma_{0}}{h^2+r^2} \exp{\big( -\frac{4r^2}{W^2(h)}}\big)\Big)        \ {\rm{d}}r\\
         &=\frac{M-2}{2\log(2)MR^2}B_{\rm{L}} \Big[ (h^2+R^2)\log\Big( \frac{\gamma_{0}}{h^2+R^2}\Big)+R^2 - h^2\log\Big( \frac{\gamma_{0}}{h^2}\Big) -\frac{2R^4}{W^2(h)}\Big].     \end{split}
    \label{eq_ASE_Central}
\end{equation}

For a single-user case, the average data rate of the VCSEL array system, $\Bar{\zeta}_{\rm{single}}^{\rm{sys}}$, is the mean of the average data rate of all beams, which can be expressed as:
\begin{equation}
    \Bar{\zeta}_{\rm{single}}^{\rm{sys}}= \frac{1}{N_{\rm{beam}}}\sum\limits_n^{N_{\rm{beam}}}\Bar{\zeta}_{\rm{single}}^n\le \frac{1}{N_{\rm{beam}}}\sum\limits_n^{N_{\rm{beam}}} \Bar{\zeta}_{\rm{single}}^1=\Bar{\zeta}_{\rm{single}}^1 = \Bar{\zeta}_{\rm{single}}^{\rm{UB}},
    \label{eq_singleUE_upperbound}
\end{equation}
where $\Bar{\zeta}_{\rm{single}}^{\rm{UB}}$ denotes the upper bound of the system average data rate in a single-user scenario. According to \eqref{eq_singleUE_upperbound}, the upper bound is the average data rate of the central beam, $\Bar{\zeta}_{\rm{single}}^1$.

\subsection{System with Multiple users}
The performance of a multi-user system is limited by two factors: noise and inter-cell interference (ICI).  When multiple UEs are served by the same beam, due to the high-directionality of each Tx in the VCSEL array system, SDMA proposed in \cite{ZheADT} will be adopted. Based on \eqref{beam_index1}, the UEs served by the same beam are grouped together. Multiple beams can serve corresponding UEs simultaneously within the same time slot as the interference between these narrow beams can be significantly mitigated. Active UEs served by the same beam share the bandwidth resource equally. Hence, the data rate of the optical SDMA system can be expressed as:
\begin{equation}
   \zeta_{\rm{multi}}^{\rm{sys}} = \sum\limits_{n=1}^{N_{\rm{beam}}} \sum\limits_{\mu=1}^{N^n_{\rm{UE}}} \zeta_{{\rm{multi}},\mu}^n = \sum\limits_{n=1}^{N_{\rm{beam}}} \sum\limits_{\mu=1}^{N^n_{\rm{UE}}} \frac{1}{N^n_{\rm{UE}}} \zeta_{\rm{single,\mu}}^n 
\end{equation}
where $\zeta_{{\rm{multi}},\mu}^n$ is the data rate of user $\mu$ served by $n$-th beam in the multi-user scenario; $N^n_{\rm{UE}}$ is the number of active UEs served by the $n$-th beam; and $\zeta_{\rm{single,\mu}}^n$ is data rate of user $\mu$ served by $n$-th beam in a single-user scenario.
For the best-case scenario, it is assumed that all the beams are utilizing different wavelengths, which should be perfectly distinguished by the optical receiver. Therefore, the system is noise limited as ICI is completely avoided and the signal to interference-plus-noise ratio (SINR) is represented as shown in \eqref{eq_snr_central_cell}. 
Hence, the average data rate of the central beam can be written as:
\begin{equation}
    \Bar{\zeta}_{\rm{multi}}^1 = \mathbb{E} \big[ \sum\limits_{\mu=1}^{N^1_{\rm{UE}}}\zeta_{{\rm{multi}},\mu}^1 \big] 
     = \sum\limits_{\mu=1}^{N^1_{\rm{UE}}} \frac{1}{N^1_{\rm{UE}}} \mathbb{E} \Big[  \zeta_{\rm{single}}^1\big(r_\mu\big) \Big]=\Bar{\zeta}_{\rm{single}}^1,
\end{equation}
where $\zeta_{{\rm{multi}},\mu}^1$ is the data rate of user $\mu$ served by the central beam, and $r_\mu$ is the horizontal distance between the user $\mu$ and the center of central cell. We can see that the average data rate of the central beam for the multi-user scenario is the same as the one for the single-user scenario. It should be noted that the average data rate of other beams is less than the average data rate of the central beam. Hence, the average data rate of the entire system for multi-user scenario will be upper bounded as follows:
\begin{equation}
    \Bar{\zeta}_{\rm{multi}}^{\rm{sys}}
    =  \sum\limits_{n=1}^{N_{\rm{beam}}}  \Bar{\zeta}_{\rm{multi}}^n  
    \le \Bar{N}_{\rm{a}} \Bar{\zeta}_{\rm{multi}}^1 = \Bar{\zeta}_{\rm{multi}}^{\rm{UB}},
    \label{eq_ASE_Multi_UE_UB}
\end{equation}
where $\Bar{N}_{\rm{a}}$ denotes the average number of active beams and $\Bar{\zeta}_{\rm{multi}}^{\rm{UB}}$ denotes the upper bound of the average data rate in the VCSEL array system. The value of $\Bar{N}_{\rm{a}}$ depends on the total number of active UEs, $N_{\rm{UE}}$, in the system as well as the total number of beams, $N_{\rm{beam}}$, on the Tx, and it can be calculated as follows \cite{OccupancyProblem}:
\begin{equation}
    \Bar{N}_{\rm{a}} = N_{\rm{beam}} - N_{\rm{beam}}(1-1/N_{\rm{beam}})^{N_{\rm{UE}}}.
\end{equation}
If the number of active UEs are infinite, $N_{\rm{UE}}=\infty$, the average number of active beams, $\Bar{N}_{\rm{a}}$, is equal to the total number of beams on the Tx. On the other hand, when there is only one active UE, $\Bar{N}_{\rm{a}}=1$.

\begin{table}[!b]
\centering
\caption{Parameters in Zemax simulation }\label{Tablezemax} 
\begin{tabular}{|c|c|c|c|c|c|}
\hline
$d_{\rm{beam}}$ & $L_{\rm{CCR}}$ & $d_{\rm{RxAP}}$ & AP height & UE height & $\theta_{\rm{FWHM}}$ \\
\hline    
12 mm & 5 mm & 5 mm & 3.5 m & 1.5 m & $4^\circ$\\
\hline
\end{tabular}
\label{Zemaxparameter}
\vspace{-20pt}
\end{table}

\section{Simulation Results}
\label{SectionSimulationResults}

\begin{figure*}[!t]
	\centering
	\begin{subfigure}[!ht]{0.5\textwidth}
		\centering
		\includegraphics[width=\columnwidth]{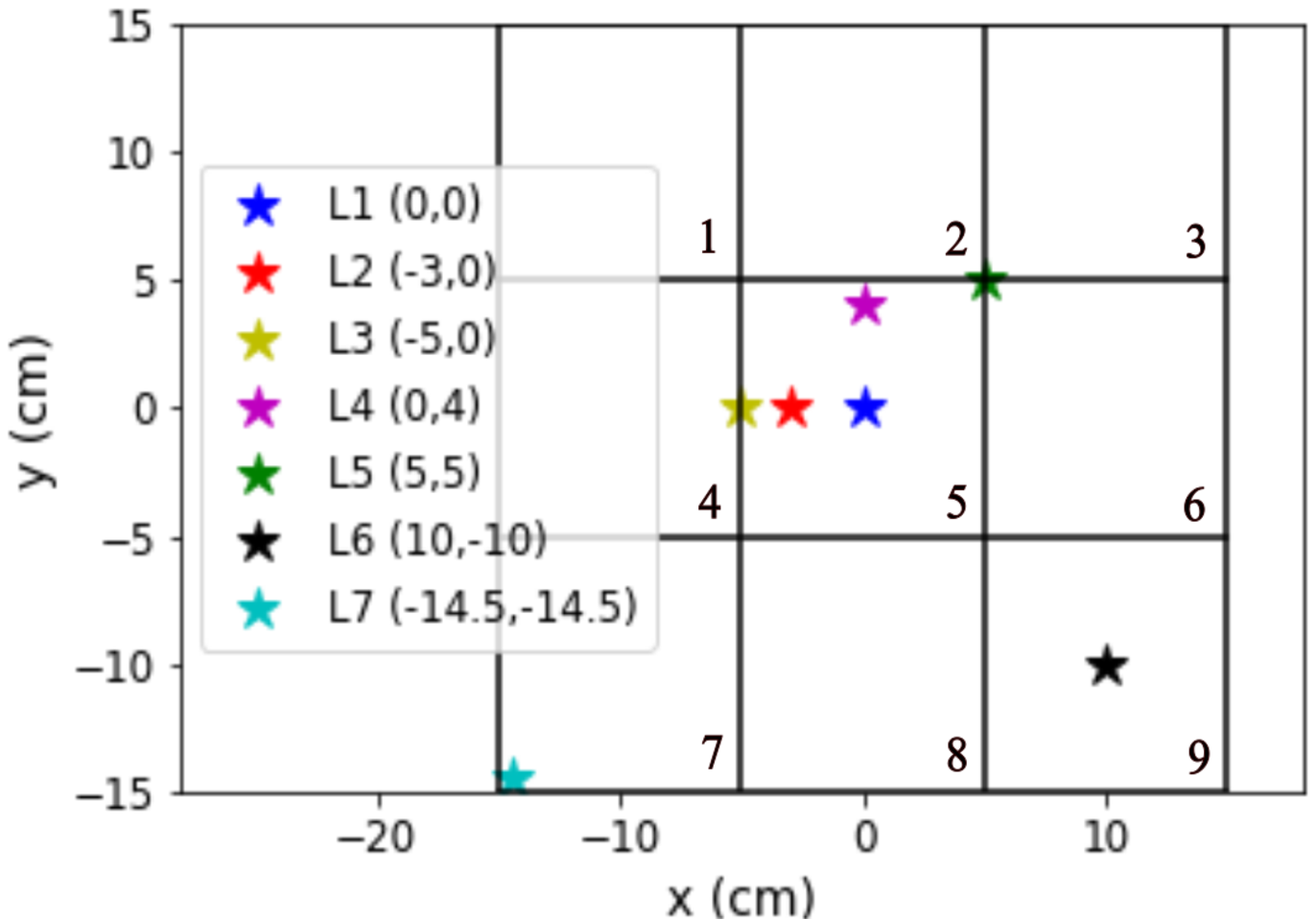}
		\caption{UE locations}
		\label{fig_UElocation}
	\end{subfigure}%
	~ 
	\begin{subfigure}[!ht]{0.5\textwidth}
		\centering
		\includegraphics[width=\columnwidth]{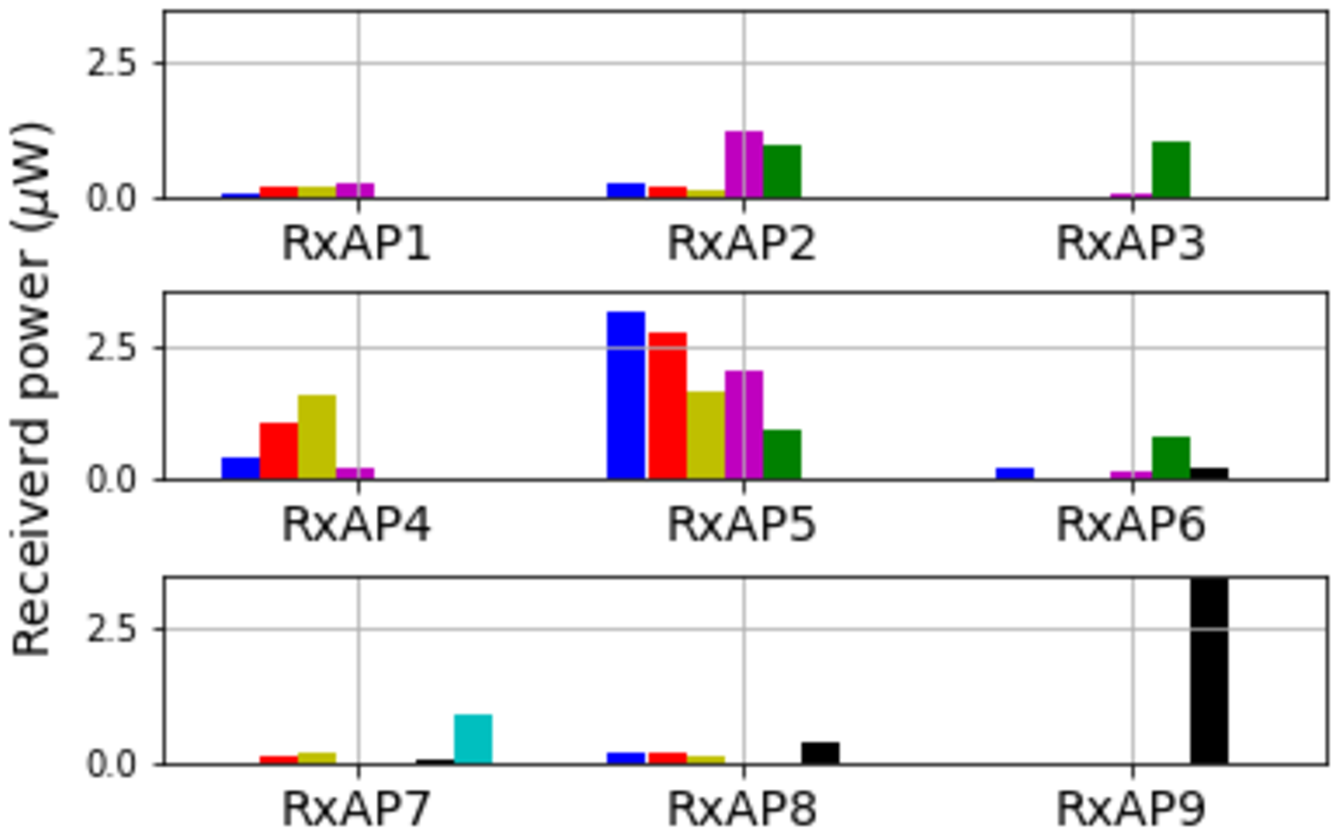}
    	\caption{Received optical power from CCR the array of RxAPs}
    	\label{fig_BarplotPowerMatrix}
	\end{subfigure}
	\caption{The UE locations and received power matrix, $N_{\rm{beam}}=9$.}
	\label{fig_powermatrix}
	\vspace{-10pt}
\end{figure*}

First of all, it should be noted that for all simulations in this study, the corresponding transmit power of beams for different values of $\theta_{\rm{FWHM}}$ are based on the eye safety regulations given in Appendix. Hence, the transmit power for $\theta_{\rm{FWHM}}$ of $2^\circ$, $4^\circ$ and $6^\circ$ are 19, 60 and 129 mW, respectively. To test the accuracy of the beam activation using CCR, the simulation with  a beam number of $N_{\rm{beam}}=9$ is carried out in Zemax and the result is presented in Fig. \ref{fig_powermatrix}. The parameters are listed in Table \ref{Zemaxparameter}.  Each beam is pointed to the center of each square cell and the cell size is \mbox{10 cm $\times$ 10 cm}. Seven UE positions are assumed as shown in \mbox{Fig. \ref{fig_UElocation}} and the received power of each PD in the array of RxAPs is shown in \mbox{Fig. \ref{fig_BarplotPowerMatrix}}. When the UE is in L1, L2 or L4, which are inside cell 5, it can be seen that the RxAP 5 receives much higher power than other RxAPs and thus beam 5 should be activated. When the UE is at L3, which is in the boundary of cell 4 and 5, the RxAP 4 and RxAP 5 receives similar power. Therefore, based on the beam activation algorithm, either beam 4 or beam 5  or both of them should be activated. When the UE is inside cell 9 such as L6, RxAP 9 has the highest power and beam 9 will be activated. L7 is located in the corner of cell 7, for this case, only the RxAP 7 collects power from the retroreflected light. These simulation results show that by using the CCR, the system can activate the corresponding beam for the UE accurately.  The IS-based VLP system proposed in \cite{VLP_zhechen} is selected as the benchmark scheme. According to \eqref{eq_time_delay} and Table \ref{table_BeamActivaiton}, the average latency, caused by the system can be calculated. Therefore, the average latency of the system using CCR is in the order of a few micro-seconds, which is significantly lower than the average positioning 
time, 44.3 ms, of the IS-based VLP system proposed in \cite{VLP_zhechen}. Hence, real-time fast tracking can be enabled by using the retroreflector. In the traditional VLP systems, if the position estimation is completed in the UE, the computation will add a high power burden to the UE. Also, in order to activate the best beam,  it requires the autonomous transmission of users' location information from  UE to APs. If the position estimation is done at the AP side, the UE needs to transmit data collected by the ISs or PDs to the AP for server-assisted computation. In conclusion, when the traditional VLP system is applied for the beam activation, it always requires a real-time backward channel from UE to APs. However, with the utilization of a retroreflector, the backward channels can be replaced by the RS reflected by the retroreflector. In addition, no additional illumination devices are required as the TS is transmitted by the VCSEL array system, which severs the dual functionality of communication and positioning. Hence, the proposed beam activation method based on a retroreflector has the advantage of low latency, low power consumption and low cost.

\begin{figure*}[!ht]
	\centering
	\begin{subfigure}[!ht]{0.45\textwidth}
		\centering
		\includegraphics[width=\columnwidth]{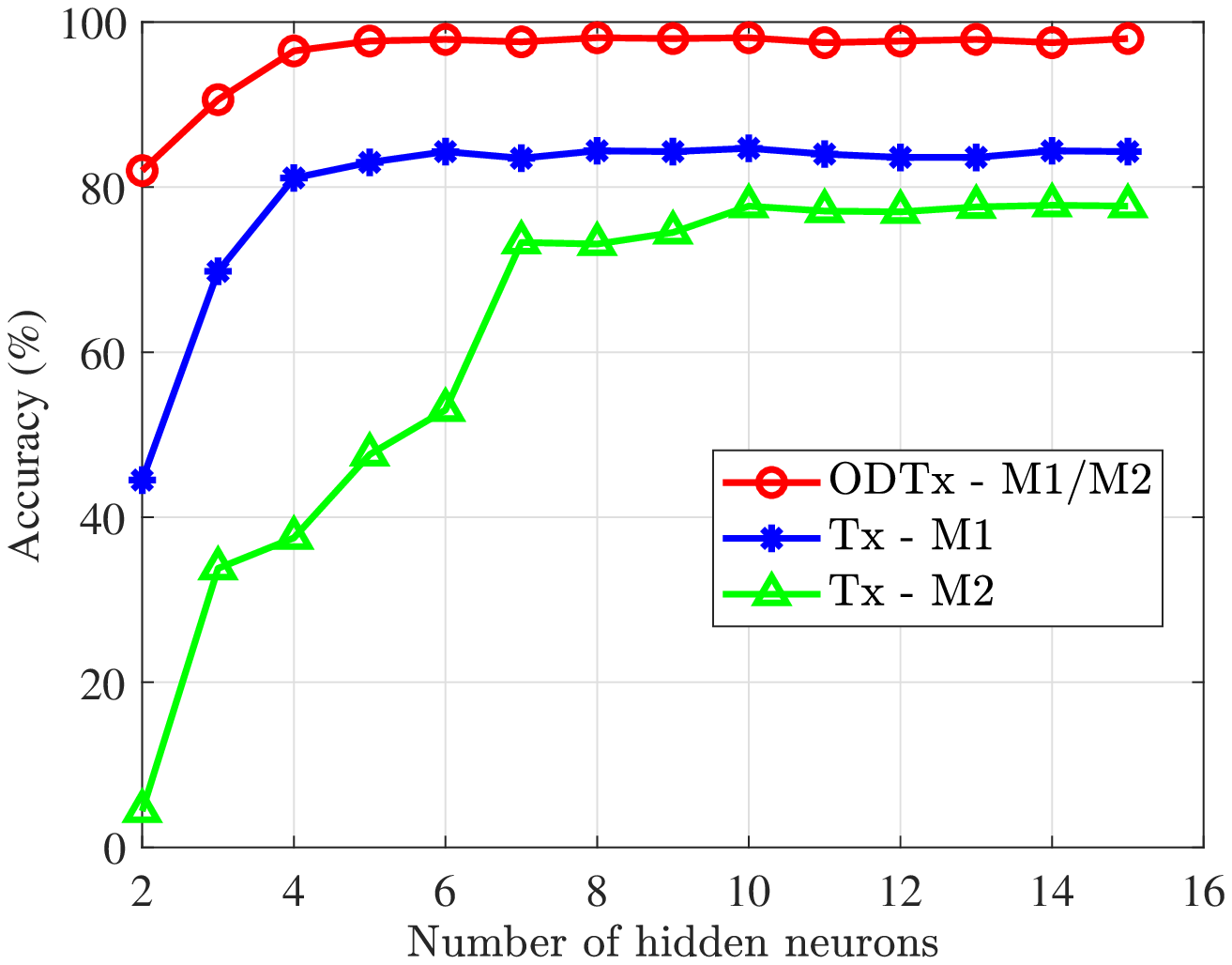}
		\caption{Accuracy against number of hidden neurons}
		\label{fig_ANN_accuracy}
	\end{subfigure}%
	~ 
	\begin{subfigure}[!ht]{0.45\textwidth}
		\centering
		\includegraphics[width=\columnwidth]{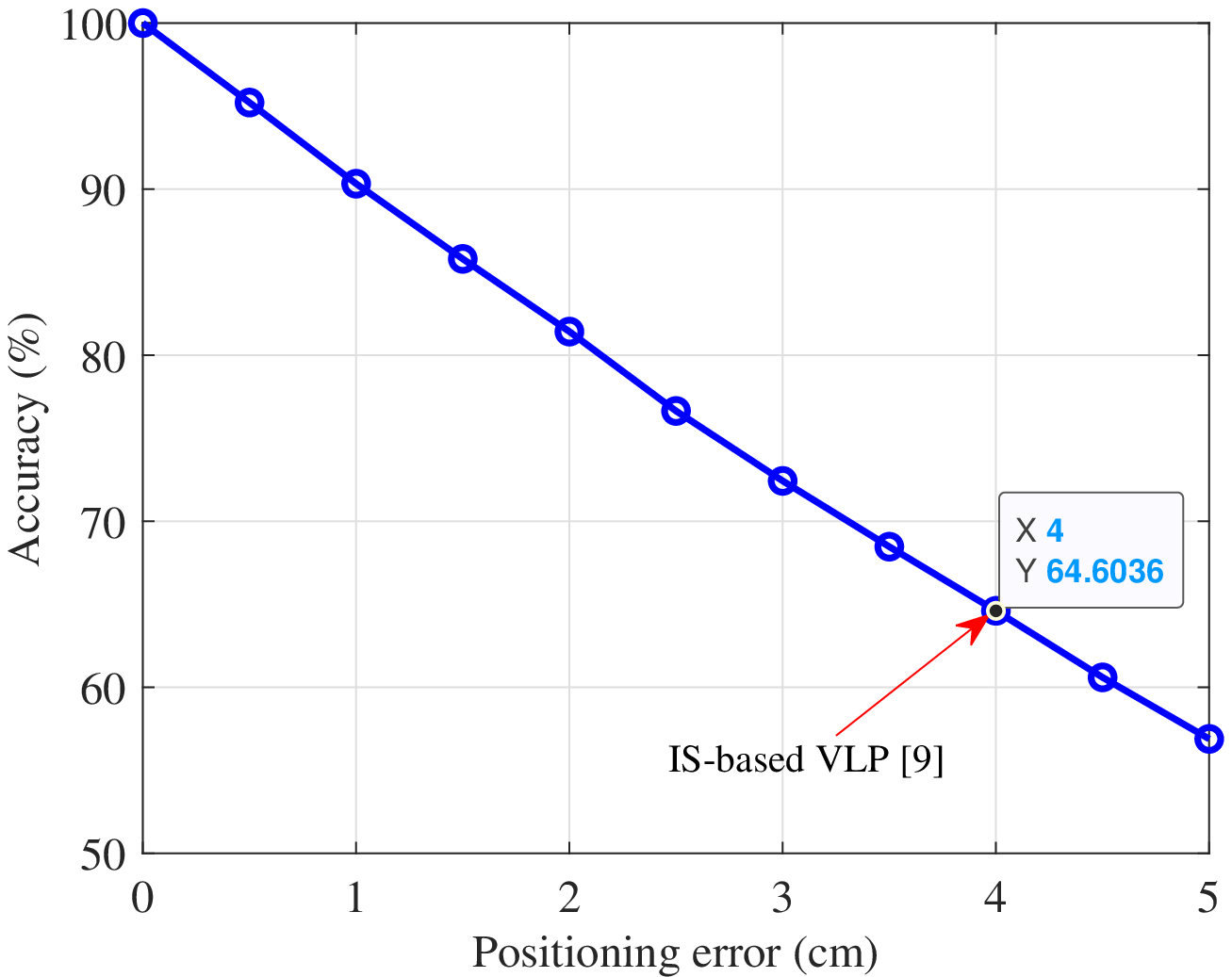}
    	\caption{Accuracy against different level of positioning error}
    	\label{fig_acc_vs_pos_error}
	\end{subfigure}
	\caption{  Beam activation accuracy}
	\label{Beam activation accuracy}
	\vspace{-10pt}
\end{figure*}

The accuracy of the beam activation is analyzed in this part. The effect of device orientation is studied and two orientation models are considered here. The first orientation model (M1) is the Gaussian model proposed in \cite{MDSorientation} for elevation angle of the UE and the second orientation model (M2) is the uniform  model, which assumes uniform distribution in all movement directions. To train the ANN, $10^5$ samples, are generated based on the analytical model in \eqref{eq_RrxUE} and \eqref{eq_PrxOD}. 80\% of the data is used for the training set and 20\% of the data is used for the test set. In a real scenario, the accuracy can be further increased by training the on-site data. As shown in Fig. \ref{fig_ANN_accuracy}, the system with ODTx achieves 98\% accuracy when $N_{\rm{hidden}}=5$. The selected benchmarks for comparison include: a) a single infrared-LED for uplink transmission and beam activation (which is called a single Tx in the rest of the paper), b) the IS-based VLP  \cite{VLP_zhechen}. Systems with a single Tx can only achieve an accuracy of 82\% and 78\% for orientation models M1 and M2, respectively.  The position error of the IS-based VLP system is around 4 cm \cite{VLP_zhechen}, which leads to a poor activation accuracy of 64.6\%, as shown in Fig. \ref{fig_ANN_accuracy}. To achieve an accuracy of more than 90\%, the positioning error is required to be less than 1 cm according to Fig. \ref{fig_acc_vs_pos_error}. In terms of the beam activation latency for the system using ODTx, the delay time, $t_{\rm{delay}}$, is mainly caused by the ANN processing time. In a laptop with Intel(R) Core(TM) i7-7700HQ CPU @ 2.8 GHz, the processing time of ANN with $N_{\rm{hidden}}=5$ for user positioning is 29.7 ms, which is lower than the 44.3 ms of the IS-based VLP systems. In addition, no additional illumination device is required for the ODTx as it serves the purpose of uplink communication and beam activation simultaneously. In conclusion, the ODTx is much more robust against random device orientation compared with the single Tx system, while it has the advantage of higher accuracy, lower latency, lower power consumption and lower cost compared with the IS-based VLP.
\begin{figure*}[!t]
	\centering
	\begin{subfigure}[!ht]{0.45\textwidth}
		\centering
		\includegraphics[width=\columnwidth]{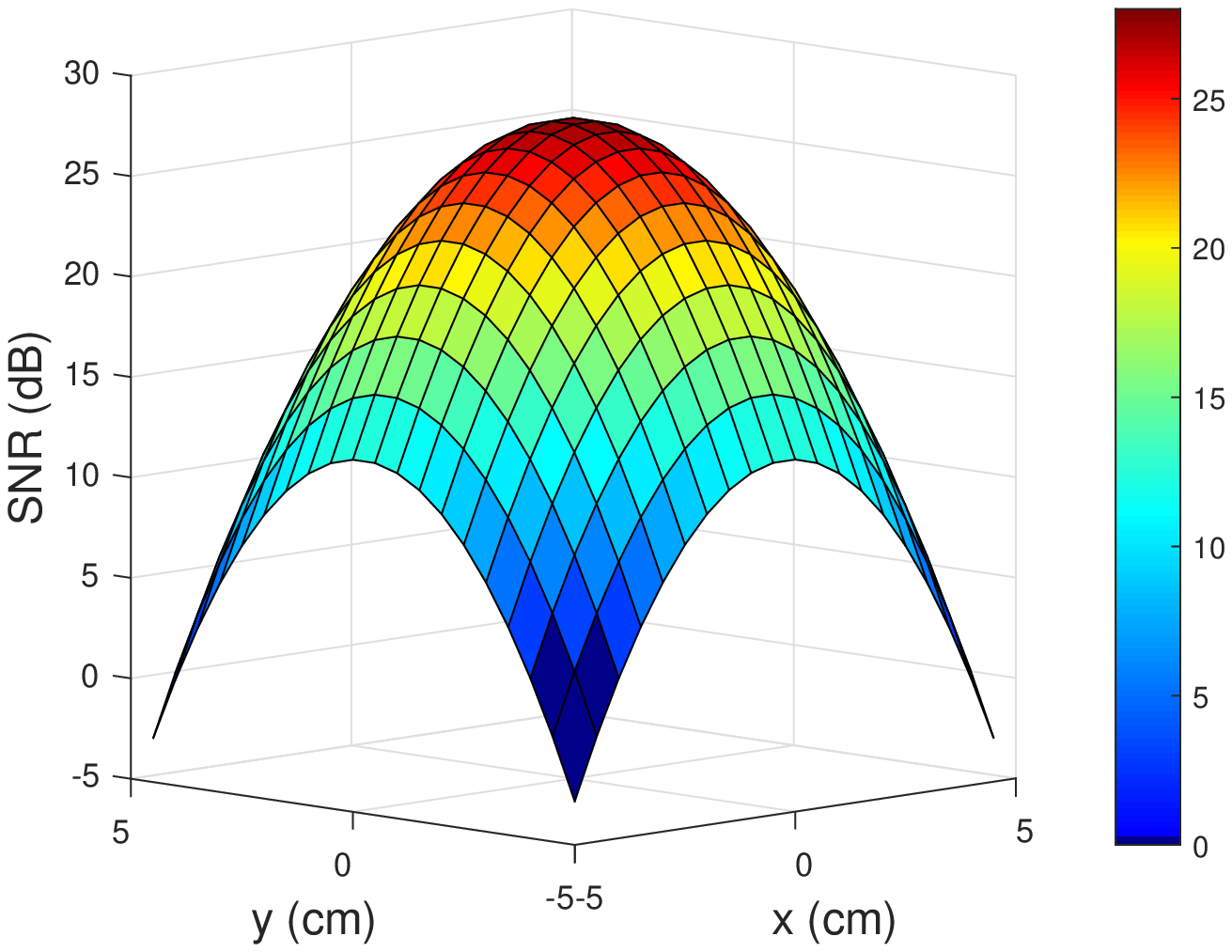}
		\caption{The central cell $(\theta_{\rm{FWHM}}=2^\circ)$.}	
		\label{fig_SNR_CentralCell1}
	\end{subfigure}%
	~
	\begin{subfigure}[!ht]{0.45\textwidth}
		\centering
		\includegraphics[width=\columnwidth]{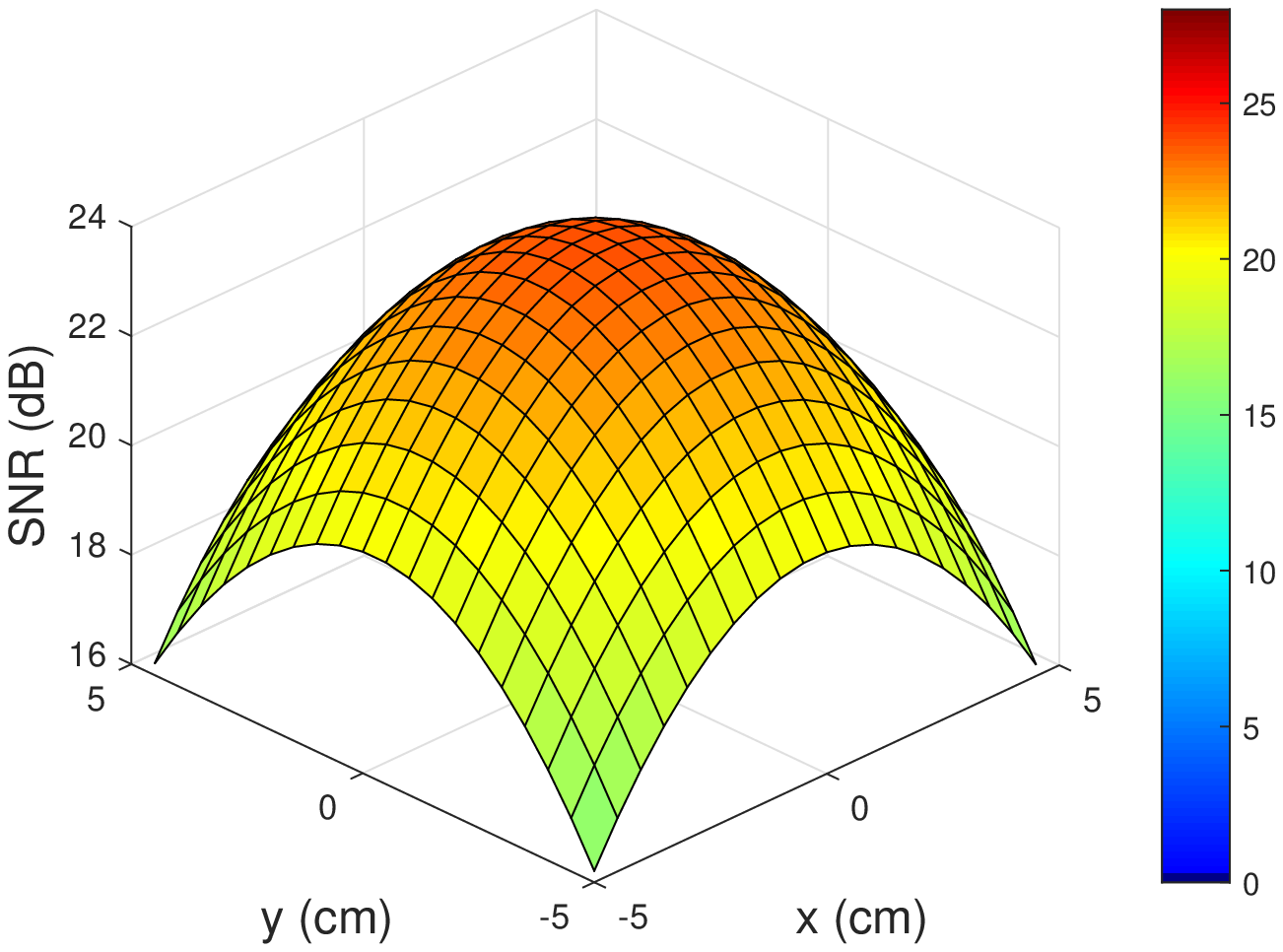}
		\caption{The central cell $(\theta_{\rm{FWHM}}=4^\circ)$.}	
        \label{fig_SNR_CentralCell2}
	\end{subfigure}
	\\
		\centering
	\begin{subfigure}[!ht]{0.45\textwidth}
		\centering
		\includegraphics[width=\columnwidth]{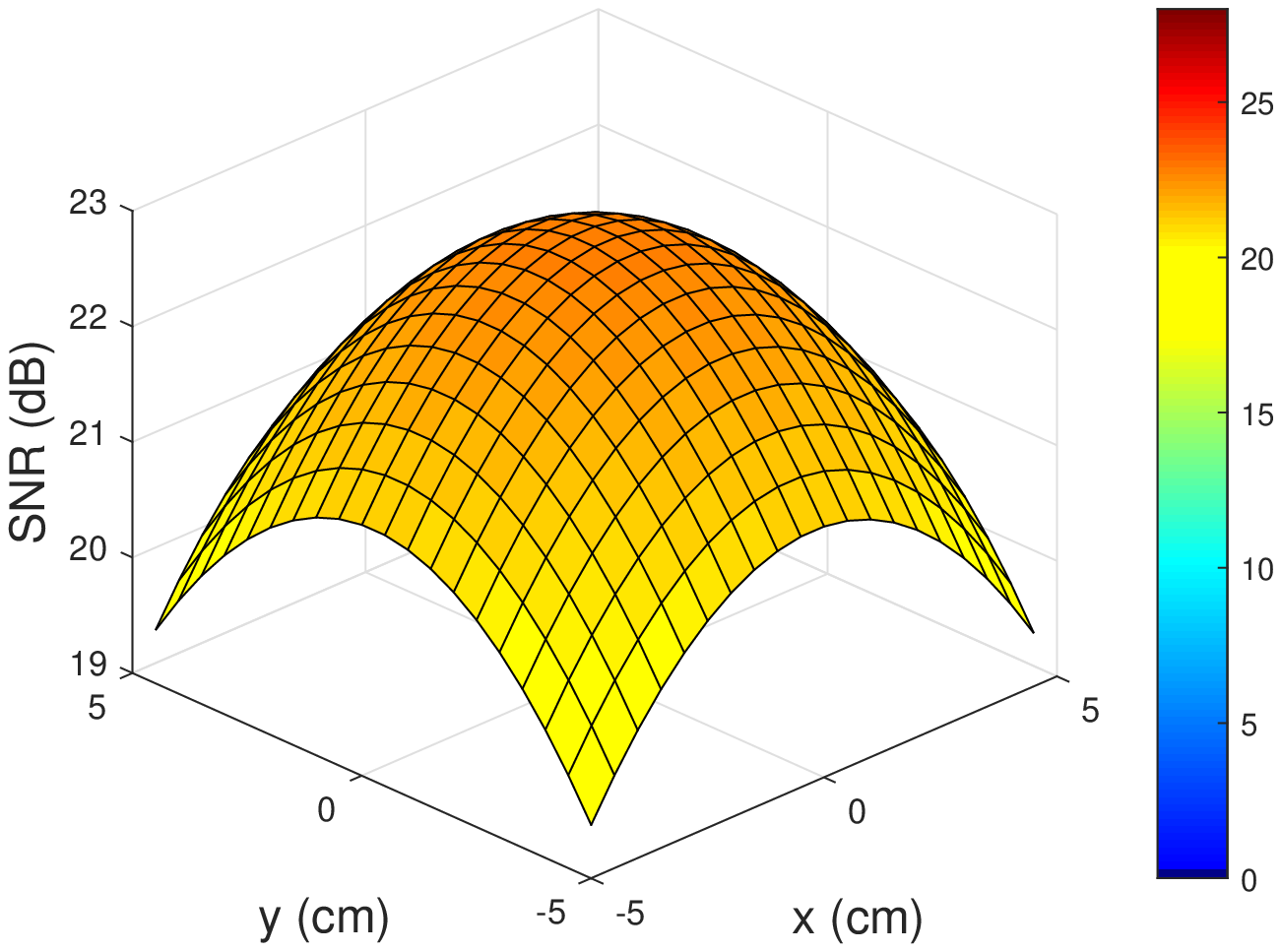}
		\caption{The central cell $(\theta_{\rm{FWHM}}=6^\circ)$.}	
		\label{fig_SNR_CentralCell3}
	\end{subfigure}%
	~
	\begin{subfigure}[!ht]{0.45\textwidth}
		\centering
		\includegraphics[width=\columnwidth]{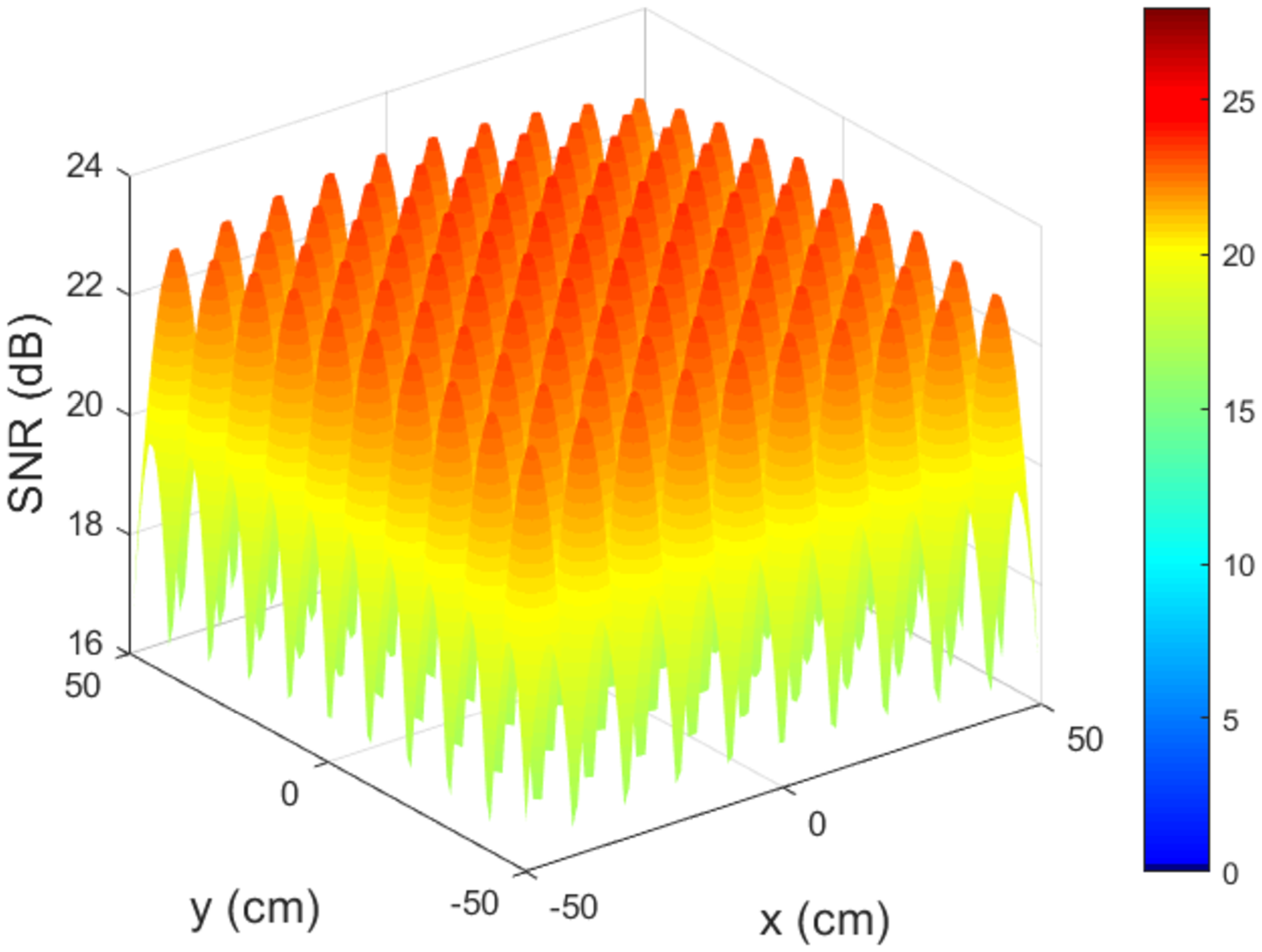}
		\caption{The VCSEL array system with $10\times10$ cells $(\theta_{\rm{FWHM}}=4^\circ)$. }	
        \label{fig_SNR_AllCell}
	\end{subfigure}
	\caption{SNR distribution in a single user system.}
    \label{fig_SNR_SingleUE}
    \vspace{-20pt}
\end{figure*}

\begin{figure*}[!t]
    \centering
        \includegraphics[width=0.55\textwidth]{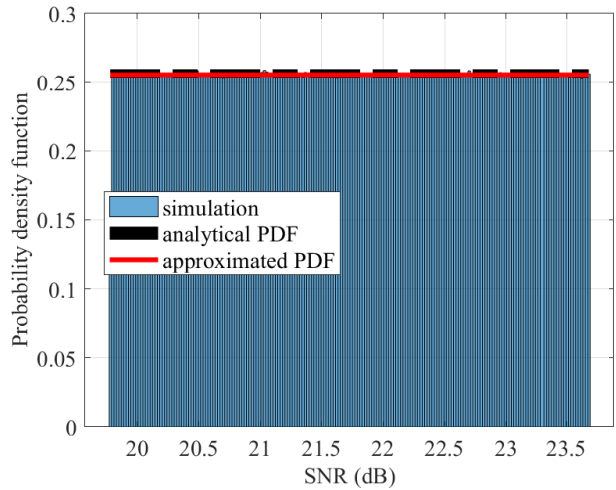}
    \caption{Probability density function of the SNR for the central beam $(\theta_{\rm{FWHM}}=4^\circ)$.}
    \label{fig_pdfsnr}  
    \vspace{-20pt}
\end{figure*}

\begin{figure*}[!ht]
    \centering
    \begin{subfigure}[!ht]{0.45\textwidth}
		\centering
        \includegraphics[width=\textwidth]{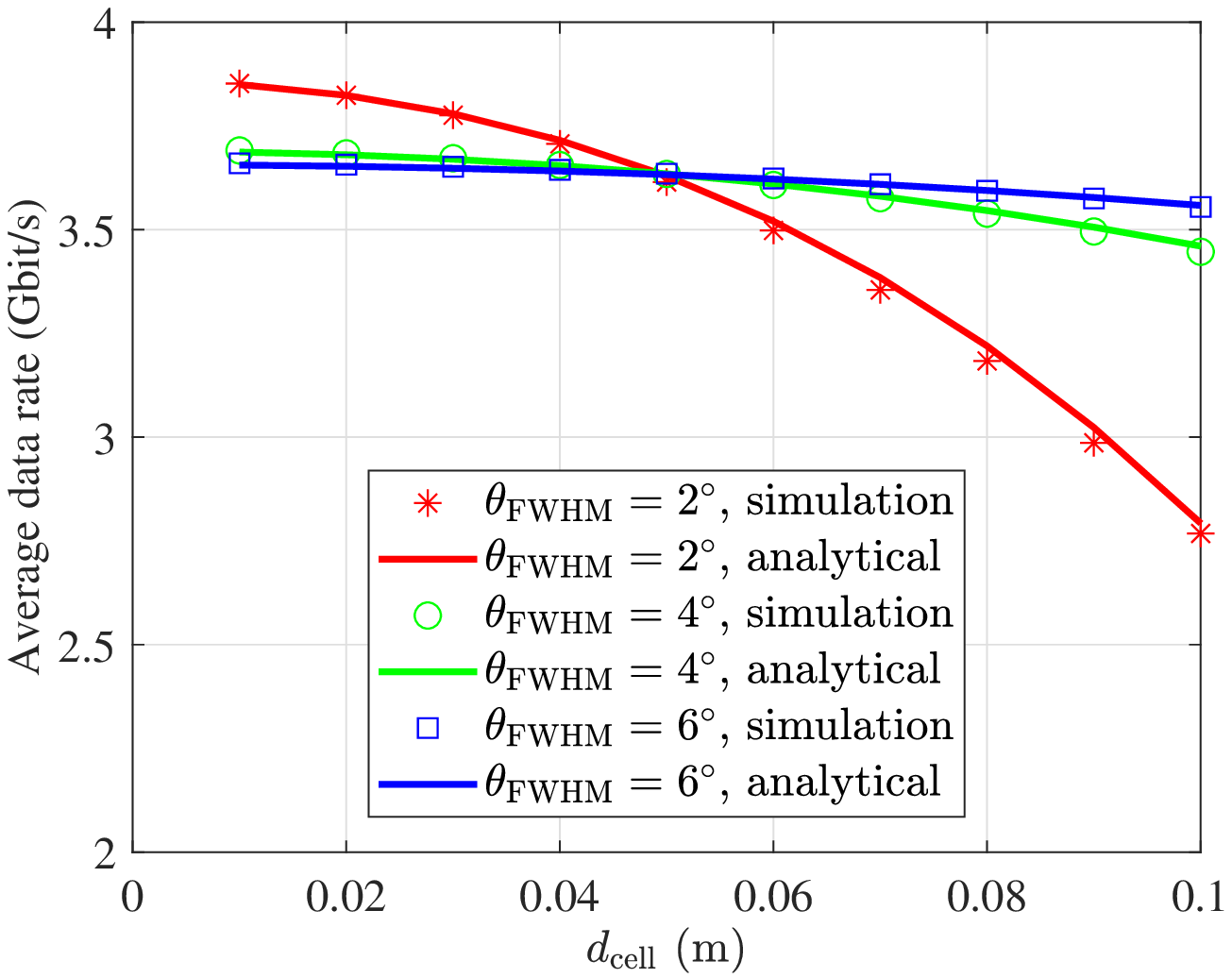}
        \caption{Average data rate for the  central cell with different cell size.}
	    \label{fig_ASE_CentralCell}
	\end{subfigure}%
    ~
    \begin{subfigure}[!ht]{0.45\textwidth}
	\centering
    \includegraphics[width=\textwidth]{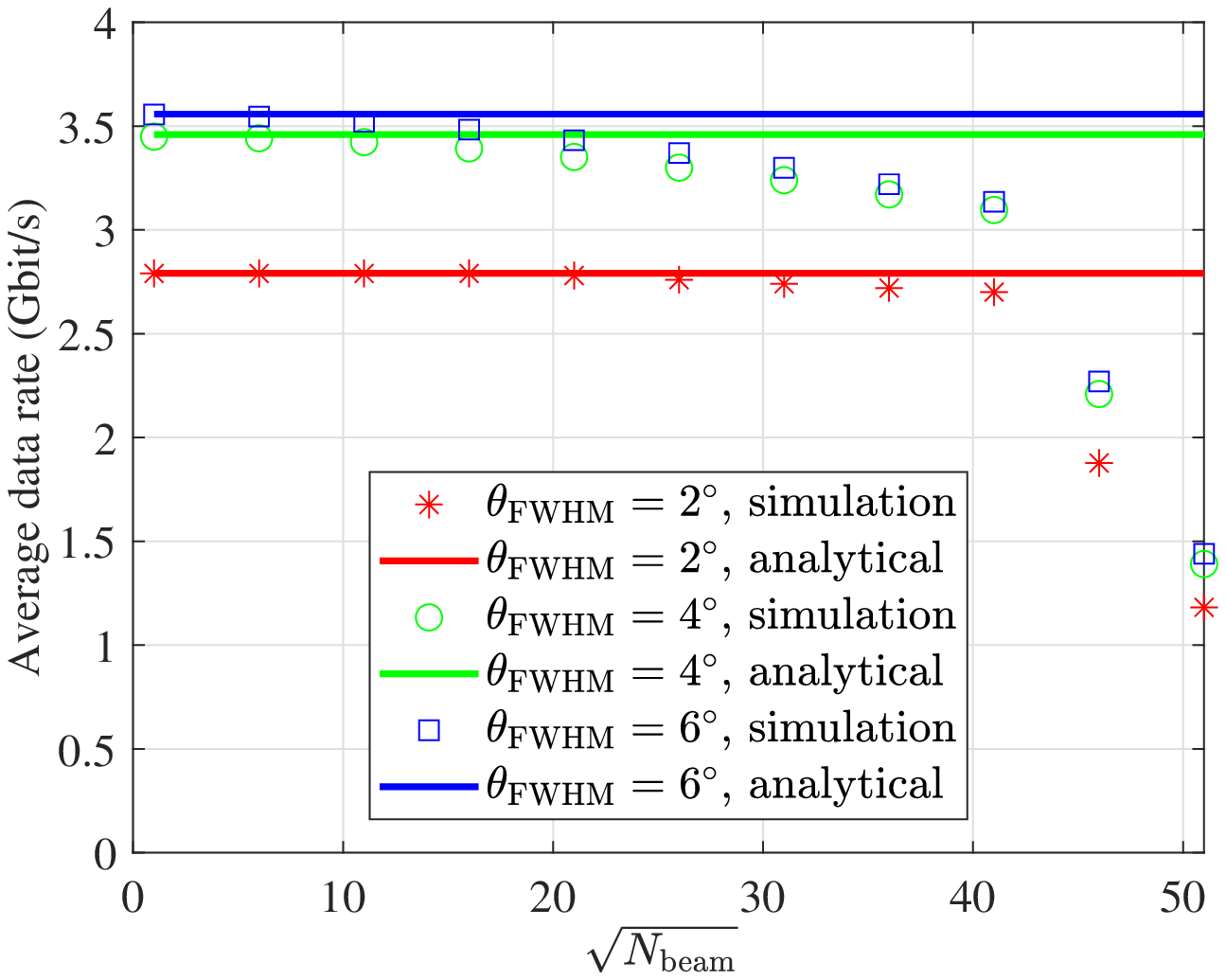}
    \caption{Average data rate for the VCSEL array system with $\sqrt{N_{\rm{beam}}}\times\sqrt{N_{\rm{beam}}}$ cells. ($d_{\rm{cell}}=10$ cm)}
	\label{fig_ASE_AllCell}
	\end{subfigure}
	\caption{Average data rate in a single user system.}
	\label{fig_ASE_SingleUE}   
	\vspace{-20pt}
\end{figure*}

The simulation results for systems with a single user scenario are shown in Fig. \ref{fig_SNR_SingleUE}-\ref{fig_ASE_SingleUE}.  In the central cell with a size of 10 cm $\times$ 10 cm, the SNR distribution is illustrated in Fig. \ref{fig_SNR_CentralCell1}-\ref{fig_SNR_CentralCell3} for different values of $\theta_{\rm{FWHM}}$. When $\theta_{\rm{FWHM}}=2^\circ$, the UE achieves the highest SNR, 27.7 dB, in the cell center. However the SNR at the cell corner is almost 0 dB. This is because the beam divergence angle is small and thus most of the power is focused around the beam center. By increasing $\theta_{\rm{FWHM}}$ from $2^\circ$ to $4^\circ$, although the transmit power increases from 19 mW to 60 mW, the peak SNR decreases to 23.7 dB since the beam is more diverged. However, the wider divergence angle leads to a great increase, 16 dB, in the SNR for the UE at the cell corner. With the further increase of $\theta_{\rm{FWHM}}$ to $6^\circ$, the transmit power doubles to 120 mW but the SNR slightly decreases to 22.7 dB. The advantage is that the UE at the cell corner can achieve a SNR of 19.4 dB. The SNR distribution of the VCSEL array system with 10 $\times$ 10 beams is plotted in Fig. \ref{fig_SNR_AllCell}. It can be seen that the distributions of all other cells have similar distributions to the central cell with a slight decrease in the SNR values. 

To verify the PDF of the SNR derived for the central beam in \eqref{eq_pdfsnr}, Monte-Carlo simulations are carried out and the histogram of SNR are plotted by the blue bars in Fig. \ref{fig_pdfsnr}. The analytical PDF in \eqref{eq_pdfsnr} is plotted by the black dash-dot line while the approximated PDF in \eqref{eq_pdfsnr_approx} is plotted by the red line. As the red line, black dash-dot line and the histogram are well matched, the PDF of the SNR for the central beam can be well approximated by the uniform distribution expressed in \eqref{eq_pdfsnr_approx}.

Fig. \ref{fig_ASE_SingleUE} shows the average data rate for the central cell with different cell size. It can be seen that for different values of $\theta_{\rm{FWHM}}$, the analytical derivation in \eqref{eq_ASE_Central} matched the Monte-Carlo simulation results. When the cell size is small, the central beam with $\theta_{\rm{FWHM}}=2^\circ$ provides higher average data rate. However, with the increase in cell size, its average data rate decreases faster than the others.  This is because with a smaller $\theta_{\rm{FWHM}}$, the beam power is more focused on the center, which leads to high power in the cell center and low power in the edge, as shown in Fig. \ref{fig_SNR_CentralCell1}.  By increasing $\theta_{\rm{FWHM}}$, the beam power is less focused. Therefore, the average date rate decreases slower when the cell size increases. When the cell size $d_{\rm{cell}}$ is 5 cm, the beam with $\theta_{\rm{FWHM}}=2^\circ$ provides the highest data rate, 3.8 Gbit/s. 
However, when the cell size $d_{\rm{cell}}$ is 10 cm, the average data rate for $\theta_{\rm{FWHM}}=4^\circ$ and $\theta_{\rm{FWHM}}=6^\circ$ are 3.4 Gbit/s and 3.5 Gbit/s, respectively, which are higher than the average data rate for $\theta_{\rm{FWHM}}=2^\circ$, 2.7 Gbit/s. Fig. \ref{fig_ASE_AllCell} demonstrates the average data rate for the VCSEL array system with $N_{\rm{beam}}$ beams and $d_{\rm{cell}}=10$ cm.  The number of beams along one axis is denoted as $\sqrt{N_{\rm{beam}}}$. It can be seen from Fig. \ref{fig_ASE_AllCell} that by varying the number of beams along one axis, the value of average data rate for the simulation results will be upper bounded by $\Bar{\zeta}_{\rm{single}}^{\rm{UB}}$ in \eqref{eq_singleUE_upperbound}.
The result indicates that, in a VCSEL array system, $\Bar{\zeta}_{\rm{single}}^{\rm{UB}}$ is actually a good approximation of the average data rate when the number of cells is small. This occurs because the distance between the central cell and the edge cell is small, the SNR distribution of other cells will be similar to the SNR distribution of the central cell as shown in Fig. \ref{fig_SNR_AllCell}. However, with the further increase of $\sqrt{N_{\rm{beam}}}$, the distance between the central cell and the edge cell increases, which will lead to higher path loss and a decrease in the average data rate. In such cases, $\Bar{\zeta}_{\rm{single}}^{\rm{UB}}$ can only be considered as the upper bound instead of a well-approximation. It should be noted that for a single UE scenario with $d_{\rm{cell}}=10$ cm, a VCSEL array system with $\theta_{\rm{FWHM}}=4^\circ$ and $6^\circ$ can provide similar performance, which is better than the system with $\theta_{\rm{FWHM}}=2^\circ$. 

\begin{figure}[!t]
    \centering
    \includegraphics[width=0.6\textwidth]{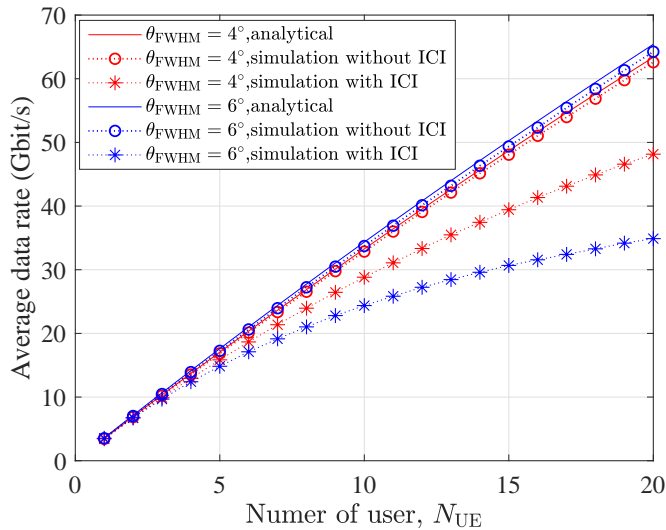}
    \caption{Average data rate for the VCSEL array system with multiple users. ($N_{\rm{beam}}=10\times10$, $d_{\rm{cell}}=10$ cm). }
    \label{fig_ASE_multiuser}
    \vspace{-20pt}
\end{figure}
The simulation for systems with multiple users are carried out. The relationship between the number of UEs, $N_{\rm{UE}}$, and the average data rate of the VCSEL array system ($N_{\rm{beam}}=100$ and $d_{\rm{cell}}=10$ cm) is shown in \mbox{Fig. \ref{fig_ASE_multiuser}}.  For the ideal case without ICI, where each beam uses a different wavelength, the analytical derivation in \eqref{eq_ASE_Multi_UE_UB} matches with the simulations without ICI for both $\theta_{\rm{FWHM}}=4^\circ$ and $6^\circ$. This also gives the upper bound of the average data rate, $\Bar{\zeta}_{\rm{multi}}^{\rm{UB}}$, for the VCSEL array system with multiple UEs. When all the beams are using the same wavelength, the ICI needs to be taken into consideration as the ICI will degrade the system performance. When the number of UEs, $N_{\rm{UE}}$, is less than 6, we can see the average data rate of the system with ICI is very close to the average data rate of the system without ICI, and thus the average data rate of the system with ICI can be well approximated by \eqref{eq_ASE_Multi_UE_UB} for small value of $N_{\rm{UE}}$. With the increase of $N_{\rm{UE}}$ from 6 to 20, the chance of having ICI rises, which widens the gap between the average data rate of simulations with ICI and the average data rate of a system without ICI. Further increases in $N_{\rm{UE}}$ from 20 will lead to a larger gap between systems with and without ICI.
It should be noted that in a VCSEL array system with multiple UEs, the system with $\theta_{\rm{FWHM}}=4^\circ$ achieves a higher data rate than the system with $\theta_{\rm{FWHM}}=6^\circ$. When the cell size is fixed, a larger divergence angle means more power will reach the adjacent cells and will be treated as interference noise by UEs in the adjacent cells. Therefore, a small divergence angle can alleviate the ICI and increase the average data rate. When $N_{\rm{UE}}=20$, a throughput of around 50 Gbit/s can be achieved by systems with $\theta_{\rm{FWHM}}=4^\circ$, while only 33 Gbit/s can be achieved for a systems with $\theta_{\rm{FWHM}}=6^\circ$. 

\begin{figure}[!t]
    \centering
    \includegraphics[width=0.6\textwidth]{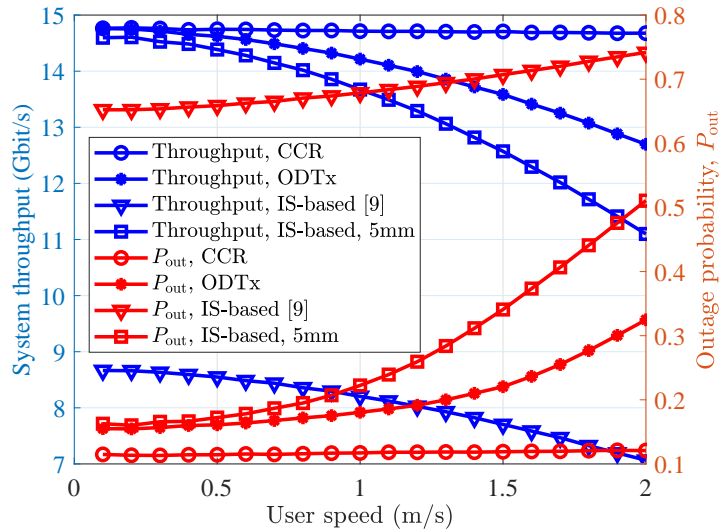}
    \caption{Evaluation of system throughput and outage probability corresponding to user speed. ($N_{\rm{beam}}=10\times10$, $d_{\rm{cell}}=10$ cm, $\theta_{\rm{FWHM}}=4^\circ$), $N_{\rm{UE}}=5$.}
    \label{fig_throughput}
    \vspace{-20pt}
\end{figure}

To calculate the system throughput, the user mobility should be considered and hence the effect of the delay caused by the beam activation should be evaluated. It is assumed that all of the users are moving randomly following the random waypoint model \cite{RandomWayPoint}. In terms of the mechanism using CCR, the effective throughput, $\mathcal{T}_{\rm{eff}}$, can be obtained based on \eqref{eq_throughput_CCR}. According to the parameters in Table \ref{table_BeamActivaiton}, it can be obtained that $\mathcal{T}_{\rm{eff}}=0.98\times\zeta_{\rm{down}}$. For the mechanism using ODTx, the latency, $t_{\rm{delay}}$, is majorly caused by the ANN processing time. Hence, the beam activated at current time will be based on the user location in $t_{\rm{delay}}$ ms ago. In a laptop with Intel(R) Core(TM) i7-7700HQ CPU @ 2.8 GHz, the processing time of ANN with $N_{\rm{hidden}}=5$ is 30 ms. The selected benchmarks for comparison include: a) the IS-based VLP proposed \cite{VLP_zhechen}, b) the IS-based VLP system in which the position error is assumed to be 5 mm. The evaluation of system throughput and outage probability corresponding to the user speed is presented in \mbox{Fig. \ref{fig_throughput}}. The outage probability, $P_{\rm{out}}$, is defined as the probability that the user throughput is less than the required threshold data rate, $R_{\rm{T}}$, where $R_{\rm{T}}$ is assumed to be 2.5 Gbit/s. When the IS-based VLP \cite{VLP_zhechen} is adopted for positioning and beam activation, due to the high positioning error and high latency, the effective system throughput is much lower than the other schemes while the outage probability is more than 60\%. When the user speed is 0.1 m/s, by decreasing the positioning error of the IS-based scheme from 3.97 cm to 5 mm, the system throughput increases substantially from 8.7 Gbit/s to 14.5 Gbit/s, which achieves similar performance as the system with CCR and the system with ODTx. However, for high-speed users, the latency, caused by beam activation and positioning, can lead to outdated location information, and the wrong beam may be activated. Therefore, when the user speed increases to 2 m/s, the system throughput of the IS-based scheme (5 mm positioning error) decreases rapidly to 11 Gbit/s. In comparison, due to the lower latency, the system throughput of the ODTx-based scheme only decreases to 12.8 Gbit/s. In terms of the CCR-based scheme, the system throughput and outage probability is independent of the user speed, which indicates the CCR mechanism has a extremely low latency and thus the effect of delay can be ignored.

\section{Conclusion}
\label{SectionConclusion}
In this study, a novel VCSEL array system, which supports high data rate, low latency and multiple user access without the requirement of expensive/complex hardware, is proposed. In addition, the choices of cell size and beam divergence angle are recommended. In order to support mobile users, two beam activation methods are proposed. The beam activation based on the CCR can achieve low power consumption and almost-zero delay, allowing real-time beam activation for high-speed users. The mechanism based on the ODTx is suitable for low-speed users and very robust to random orientation. The proposed methods exhibits significantly superior data rate and outage probability performance than the benchmark schemes. For a single UE scenario, regarding the central beam, the PDF of the SNR in dB can be considered as a uniform distribution. In addition, the analytical derivations for the average data rate of the central beam and its upper bound are presented and proved by simulations. This study may have great potential in guiding the designs in future experiment studies and performance analysis on the VCSEL array system. The research area is still broadly open for research and could be extended in many interesting directions, such as utilizing the spatial and multiplexing gain, designing the omnidirectional CCR and optimizing the resource management in a multi-VCSEL-array network.

\begin{figure*}[!ht]
    \centering
    \begin{subfigure}[!ht]{0.45\textwidth}
		\centering
        \includegraphics[width=\columnwidth]{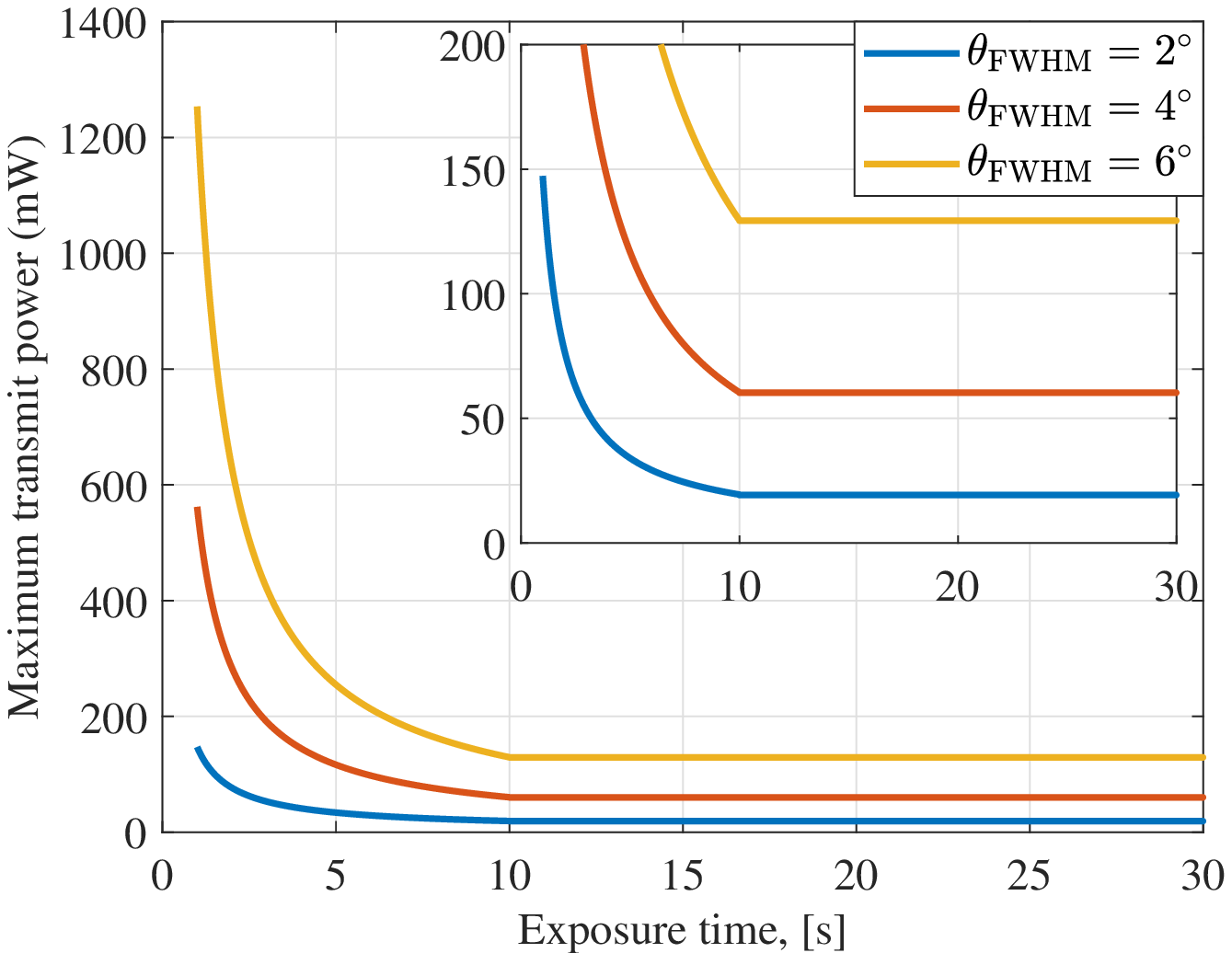}
        \caption{1550 nm}
	\end{subfigure}%
    ~
    \begin{subfigure}[!ht]{0.45\textwidth}
	\centering
    \includegraphics[width=\columnwidth]{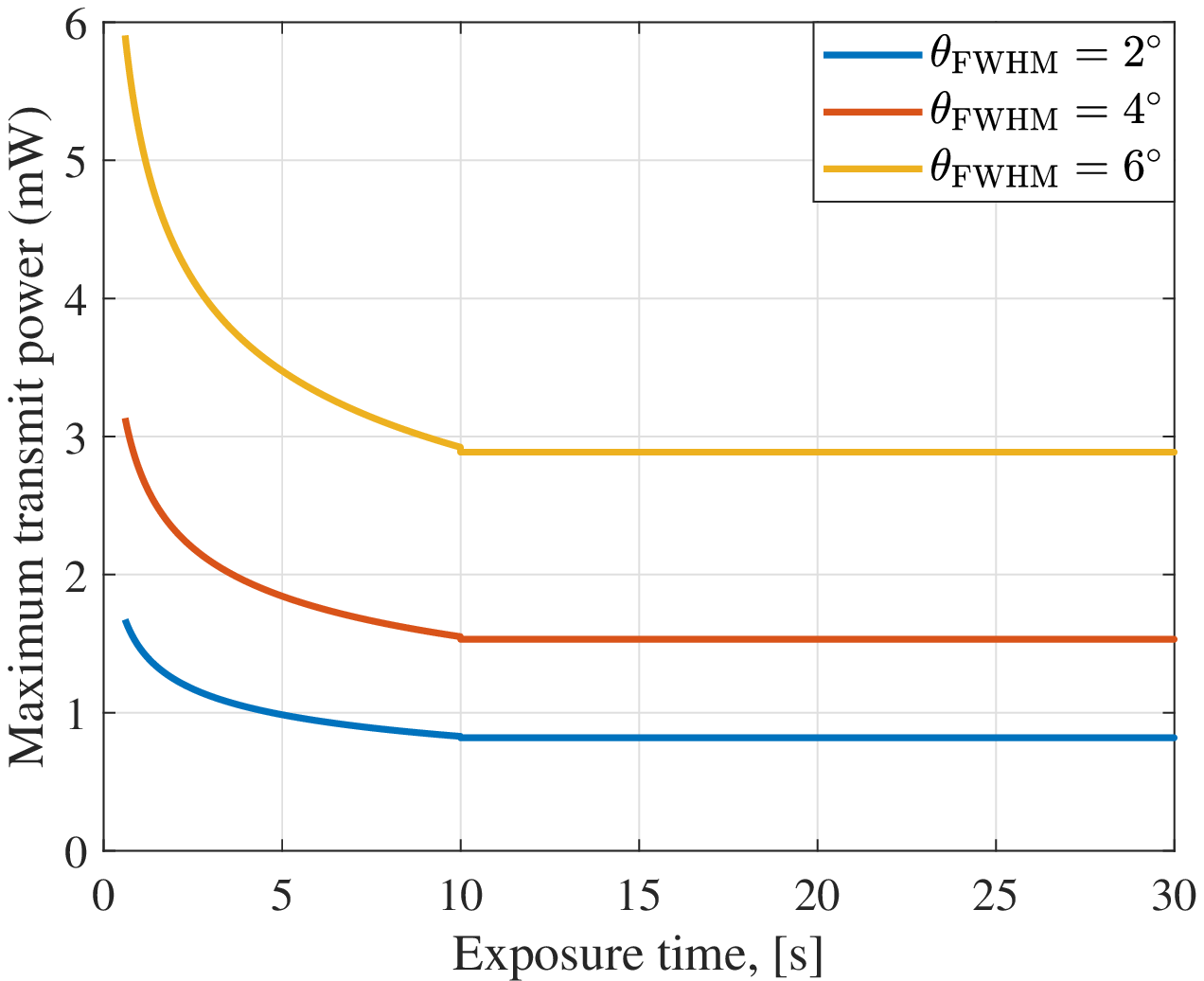}
	\caption{850 nm}
	\end{subfigure}
    \caption{Maximum transmit power considering eye safety}
\label{fig_eyesafety}
\vspace{-20pt}
\end{figure*}

\appendix
\label{appendix_eyesafety}
When laser diodes or VCSEls are deployed, it is necessary to consider eye safety, which will limit the value of the transmit power, $P_{\rm{tx,opt}}$. In this section, the maximum allowable transmit optical power $P_{\rm{tx,max}}$ will be calculated based on regulations in the eye safety standard \cite{BSEN80625}. The Gaussian beam intensity at the distance $z$ is:
\begin{equation}
    I(r,z) =\frac{2P_{\rm{tx,opt}}}{\pi W^2(z)} \exp{ \Big(-\frac{2r^2}{W^2(z)}} \Big).
\end{equation}
The aperture diameter of cornea is denoted as $d_{\rm{a}}$. The exposure level of the cornea at distance $z$ can be calculated as:
\begin{equation}
    \begin{split}
    E_{\rm{exp}}(z)&=\frac{1}{\pi(d_{\rm{a}}/2)^2} \int^{d_{\rm{a}}/2}_0 I(r,z) 2\pi r \ {\rm{d}}r \\
    &= \frac{1}{\pi(d_{\rm{a}}/2)^2} \int^{d_{\rm{a}}/2}_0 \frac{2P_{\rm{tx,opt}}}{\pi W^2(z)} \exp{ \Big(-\frac{2r^2}{W^2(z)}} \Big) 2\pi r \ {\rm{d}}r \\ 
    &= \frac{P_{\rm{tx,opt}}}{\pi(d_{\rm{a}}/2)^2} \Big(1-\exp(-\frac{d_{\rm{a}}^2}{2W^2\big(z)}\big)\Big)
    \end{split}.
\end{equation}
The level of radiation to which persons may be exposed without suffering adverse effects is called the maximum permissible exposure (MPE). Exposures above the MPE will lead to eye injuries. Instead of actually performing an MPE analysis for all locations along the beam, the exposure level at the most hazardous position (MHP), $z_{\rm{mph}}$, can be calculated and compared with the MPE. If the exposure level at the MHP, $E_{\rm{exp}}(z_{\rm{mph}})$, is less then the MPE, then the MPE is not exceeded anywhere else in the beam \cite{LaserSafetyHenderson}. Hence, when considering eye safety, $E_{\rm{exp}}(z_{\rm{mph}})\leq E_{\rm{mpe}}(t_{\rm{exp}})$, where $t_{\rm{exp}}$ represents the exposure duration and  $E_{\rm{mpe}}(t_{\rm{exp}})$ represents the MPE value corresponding to $t_{\rm{exp}}$. Therefore, the maximum allowable transmit optical power $P_{\rm{tx,max}}$ can be written as:
\begin{equation}
    P_{\rm{tx,max}}(t_{\rm{exp}}) = \frac{\pi d_{\rm{a}}^2 E_{\rm{mpe}}(t_{\rm{exp}})}{4\Big(1-\exp(-\frac{d_{\rm{a}}^2}{2W^2(z_{\rm{mph}})}\big)\Big)}.
\end{equation}
When the angular subtense of the apparent source is less than 1.5 mrad, the MPH, $z_{\rm{mph}}$, is \mbox{10 cm}. The MPE value $E_{\rm{mpe}}(t_{\rm{exp}})$ can be calculated based on parameters in Table \ref{table_eye_safety} \cite{BSEN80625}. The maximum allowable transmit optical power, $P_{\rm{tx,max}}$ is plotted against the exposure duration $t_{\rm{exp}}$ in Fig. \ref{fig_eyesafety} for different wavelengths and different $\theta_{\rm{FWHM}}$.
The $P_{\rm{tx,max}}$ for $\lambda=1550$ nm is much larger than the $P_{\rm{tx,max}}$ for $\lambda=850$ nm so that $\lambda=1550$ nm is chosen in this study for higher data rate communication. The maximum allowable transmitted power, $P_{\rm{tx,max}}$, for $\theta_{\rm{FWHM}}=2^\circ$, $4^\circ$, $6^\circ$ are 20 mW, 60 mW, and 130 mW, respectively.
\begin{table}[!ht]
\centering
\caption{Parameters for eye safety }
\label{table_eye_safety} 
\begin{tabular}{|c|c|c|c|c|}
\hline    
 & \multicolumn{2}{c|}{$\lambda=1550$ nm} & \multicolumn{2}{c|}{$\lambda=850$ nm} \\
\hline
Exposure duration, $t_{\rm{exp}}$ (s) & 0.35 to 10 & 10 to $10^3$ & $10^{-3}$ to 10 & 10 to $10^3$\\
\hline
MPE, $E_{\rm{mpe}}$ ($\rm{{W}}\cdot {\rm{m}}^{-2}$) & $10^4/t_{\rm{exp}}$ & 1000 & $18t_{\rm{exp}}^{0.75}{\rm{C}}_4/t_{\rm{exp}}$ &10${\rm{C}}_4{\rm{C}}_7$ \\
\hline
Aperture stop (mm) $d_{\rm{a}}$ &  $1.5t_{\rm{exp}}^{3/8}$ & $3.5$ & 7 & 7 \\
\hline
${\rm{C}}_4$ & - & -  & $10^{0.002(\lambda-700)}$ & $10^{0.002(\lambda-700)}$ \\
\hline
${\rm{C}}_7$ & - & -  & $1$ & $1$ \\
\hline    
\end{tabular}
\vspace{-20pt}
\end{table}

\bibliography{reference}

\end{document}